\documentclass[9pt]{revtex4}
\usepackage{hyperref}
\usepackage{datetime}
\usepackage{amsmath,amssymb,amsfonts,mathrsfs,caption,epstopdf,mathptmx,hyperref}
\usepackage{mdframed}
\usepackage{cases}
\usepackage{color}
\usepackage{graphicx,subcaption,float,wrapfig}
\usepackage{overpic}
\usepackage[usenames, dvipsnames]{xcolor}
\usepackage{multirow}
\usepackage{dcolumn}% Align table columns on decimal point
\usepackage{bm}% bold math

\newtheorem{proposition}{Proposition}[section]

\def\P{\mathbf{P}}
\def\Q{\mathbf{Q}}
\def\I{\mathbf{I}}

\def\n{\mathbf{n}}
\def\m{\mathbf{m}}
\def\p{\mathbf{p}}
\def\r{\mathbf{r}}
\def\x{\mathbf{x}}

\def\z{\mathbf{z}}
\def\xhat{\hat{\mathbf{x}}}
\def\yhat{\hat{\mathbf{y}}}
\def\zhat{\hat{\mathbf{z}}}
\def\div{\mathrm{div}}
\def\tr{\mathrm{tr}}

\begin{document}
\title{Pattern Formation for Nematic Liquid Crystals-Modelling, Analysis, and Applications}
\author{Yucen Han$^1$ and Apala Majumdar$^1$}
\address{$^1$Department of Mathematics and Statistics, University of Strathclyde, G1 1XQ, UK.}

\begin{abstract}
We summarise some recent results on solution landscapes for two-dimensional (2D) problems in the Landau--de Gennes theory for nematic liquid crystals. We study energy-minimizing and non energy-minimizing solutions of the Euler--Lagrange equations associated with a reduced Landau-de Gennes free energy on 2D domains with Dirichlet tangent boundary conditions. We review results on the multiplicity and regularity of solutions in distinguished asymptotic limits, using variational methods, methods from the theory of nonlinear partial differential equations, combinatorial arguments and scientific computation. The results beautifully canvass the competing effects of geometry (shape, size and symmetry), material anisotropy, and the symmetry of the model itself, illustrating the tremendous possibilities for exotic ordering transitions in 2D frameworks.
\end{abstract}
% PACS 
\maketitle
\section{Introduction}\label{sec:intro} 
Liquid crystals are partially ordered materials, intermediate between conventional solid and liquid phases.
They will typically combine fluidity with some translational or orientational order characteristic of a solid. 
Liquid crystals were accidentally discovered by Friedrich Reinitzer, an Austrian plant physiologist, when he was experimenting with cholesteryl benzoate \cite{reinitzer1888beitrage}. The unique physical, optical and rheological properties of liquid crystals were gradually unveiled with time and today, we know that liquid crystals are ubiquitous in daily life, e.g. in some clays, soap, the human DNA, cell membranes, polymers, elastomers and the list keeps growing \cite{lagerwall2016introduction}. %and even in DNA, cell membrane.
Liquid crystals can be classified as: nematic liquid crystals, cholesteric liquid crystals and smectic liquid crystals \cite{friedel1922etats}. The simplest phase is the nematic liquid crystal (NLC) phase, for which the constituent rod-like molecules exhibit long-range orientational order, with no positional order. In the cholesteric phase, the molecules naturally twist following a helical pattern and one can view the NLC phase as a special cholesteric with no twist. In the smectic phase, the molecules arrange themselves in layers and there is orientational order within the layers, and the layers can slide past each other. In this review, we focus on the mathematical modelling, analysis and simulations of NLCs in confinement, illustrating the plethora of exotic possibilities and how mathematics can be used to predict, tune and select the properties of confined NLC systems.

NLCs are the most widely used liquid crystals, perhaps because of their relative simplicity. NLC molecules are typically asymmetric in shape, e.g. rod-shaped, disc-shaped, banana-shaped or bent-core  \cite{dg}. These asymmetric NLC molecules move freely but tend to align along certain locally preferred directions, referred to as \emph{nematic directors} in the literature \cite{dg}. Consequently, NLC phases exhibit long-range orientational order and naturally have direction-dependent responses to incident light, external electric field or magnetic fields, temperature and mechanical stresses. Consequently, NLCs are anisotropic with directional physical properties such as the NLC dielectric anisotropy, magnetic susceptibility, and the optical refractive indices \cite{palffy2007orientationally}. In particular, the anisotropic NLC response to light and electric fields have made NLCs the working material of choice for the multi-billion dollar liquid crystal display (LCDs) industry  \cite{bahadur1990liquid}. NLCs have been widely used in electric billboards, TVs, laptops, calculators, and watches and a range of opto-electric devices. In recent years, there has been unprecedented interest in using NLCs for the design of new meta-materials, bio-materials, composite materials, all of which render new possibilities for sensors, photonics, actuators, artificial intelligence and diagnostics \cite{lagerwall2012new, jampani2019liquid}. %Subsequently, nematic liquid crystals has aroused widespread interest from scientists.

Mathematics can play a crucial role in predicting, manipulating and even designing tailor-made NLC systems. NLC systems can be mathematically modelled at different levels, ranging from fully molecular approaches, to mean-field approaches such as Onsager theory, Maier--Saupe theory \cite{onsager1949effects,maier1958einfache} to fully continuum approaches such as the Oseen--Frank theory, the Ericksen--Leslie theory and the celebrated Landau--de Gennes (LdG) theory  \cite{dg,Ericksen1990,oseen1933theory, dg}. We focus on continuum approaches wherein we do not focus on microscopic details or microscopic interactions, assuming that macroscopic properties of interest vary slowly on microscopic length scales. In the continuum approach, the NLC state is described by a macroscopic order parameter, that is an averaged measure of the degree of nematic orientational order. The physically observable states are modelled by local or global minimizers of an appropriately defined free NLC energy, which typically depends on the NLC order parameter, its gradient and various material-dependent and temperature-dependent phenomenological constants. Mathematically, this naturally raises highly non-trivial questions in the calculus of variations, singular perturbation theory, homogenization theory and algebraic topology. The critical points of the NLC free energy are solutions (in an appropriately defined sense) of a system of nonlinear, coupled partial differential equations -  the Euler-Lagrange equations, with different types of boundary conditions - Dirichlet, Neumann, Robin etc. for the NLC order parameter. Of particular interest are the multiplicity and regularity of solutions, and how this depends on the structure of the model, the phenomenological model parameters, the symmetry of the domain and the boundary frustration. Given that the variational problems are typically nonlinear and non-convex, there are multiple solutions of the Euler-Lagrange solutions, some energy-minimizing and some non energy-minimizing solutions \cite{majumdar2010landau}. The non-minimizing solutions play a crucial role in the selection of energy minimizers and switching mechanisms in NLC systems with multiple energy minimizers. Regarding regularity, NLC defects are interpreted as a localised region of reduced NLC orientational order, which could be induced by temperature changes or by discontinuities in the nematic directors \cite{han+majumdar+harris+zhang+2021, majumdar2010landau, majumdar2010equilibrium}. NLC defects are a fundamental optical signature of NLCs in confinement \cite{kleman1989defects, lagerwall2016introduction}. NLC defects play a crucial role in multiplicity of solutions, the solution properties and ultimately, structural transitions often proceed via the creation and annihilation of defects \cite{kusumaatmaja2015free}. There are open mathematical questions regarding the mathematical definition of a defect, and how the defect set depends on the nature of the partial order, the mathematical model and the physical variables. Collectively, liquid crystals are a fascinating playground for mechanics, geometry, modelling and analysis to drive a new revolution in mathematics-driven interactive materials science, for sweeping interdisciplinary and practical advances.

In this review, we focus on multistable two-dimensional NLC systems, driven by recent advances in new generations of bistable LCDs \cite{jones2017defects,tsakonas2007multistable}, micropatterned surfaces \cite{kim2002tristable}, and also in 3D printing \cite{gantenbein2018three}. Multistable systems can support multiple stable nematic equilibria without any external applied fields, ideally with distinct optical and physical properties, offering multiple modes of functionality. For example, in a bistable LCD, the bright (transparent) state and the dark (opaque) states are stable without any external electric fields, so that power is only needed to switch between states or to refresh the image, but not to maintain a static image  \cite{stewart2019static}. 
Therefore, bistable LCDs are efficient, low-cost displays with enhanced optical properties. Some of the results reviewed in this article are motivated by the planar bistable LC device reported in \cite{tsakonas2007multistable}. This planar device comprises a periodic array of square or rectangular NLC-filled wells, typically on the micron-scale such that the well height is much smaller than the cross-section dimensions. Hence, it is reasonable to assume that the NLC structural profile is invariant along the height of the well, and it suffices to model planar profiles in the well square cross-section. The well surfaces are treated to induce tangential or planar anchoring, so that the NLC molecules lie in the plane of the well surfaces and tangent to the well edges. There is a natural mismatch in the nematic directors at the square vertices, leading to interesting and multiple possibilities for stable NLC configurations.
%Some of our reasearches are for thin 3D systems, which is three dimensional but the height of the well is very small compared to the cross-section of the dimensions. For example, the domain of the planar bistable LC device is actually a square well \cite{tsakonas2007multistable}. Thin systems can be well-modelled in two dimensions and studied by reduced or attractable approaches.
%And from the modelling aspective, it is just assuming the 3D system is invariant in z-axis. Then the 2D domain of bistable LC device is a square.
%Actually, 2D systems are physically relevant. They come across in micropaterned surfaces \cite{kim2002tristable}, and also in 3D printing \cite{gantenbein2018three}.

Indeed, this relatively simple geometry is actually experimentally reported to be bistable \cite{tsakonas2007multistable}. There are at least two experimentally reported stable states --- the \emph{diagonal} state for the nematic director is roughly along a square diagonal, and the \emph{rotated} state for which the nematic director rotates by 180 degrees between a pair of opposite edges \cite{gantenbein2018three}.
Both states have long-term stability 
and somewhat contrasting optical properties, without external electric fields. We use this example of a multistable system as a benchmark example, and in this review, we address natural questions such as - what happens if we replace the square with a regular or asymmetric 2D polygon, what are the effects of material anisotropy on multistability and crucially, can we mathematically model solution landscapes in reduced 2D frameworks and study the connectivity of non-energy minimizing solutions to energy-minimizing solutions? The transition pathways between the diagonal and rotated states have been studied in  \cite{kusumaatmaja2015free}, but a systematic study of the non energy-minimizing critical points is largely open.

% Paper structure
The review paper is organized as follows. The LdG theory is reviewed in Section \ref{sec:theory}. In Section~$3$, we review some known results on NLC solution landscapes for square domains as a benchmark example. In Section \ref{sec:pol}, we summarize the results in \cite{han2020pol} to illustrate multistability for NLCs in 2D polygons and the effects of geometry. In Section \ref{sec:rec} and \ref{sec:ani}, we summarize the results reported in \cite{fang2019solution} and \cite{han+majumdar+harris+zhang+2021} to elucidate the effects of geometrical asymmetry (by taking the rectangle as an example) and the effects of elastic anisotropy on NLC solution landscapes on square domains. The last leg of the review concerns unstable saddle points (non-minimizing solutions) and transition pathways for NLCs, as reported in \cite{han2020SL} in Section \ref{sec:SL}. Some conclusions are discussed in Section \ref{sec:conclusion}. In the supplement, Section \ref{sec:num}, we summarize the numerical methods used for solving the complex system of LdG Euler-Lagrange equations and for computing the non-trivial NLC solution landscapes.

\section{The Landau--de Gennes theory}\label{sec:theory}

The Landau--de Gennes (LdG) theory is perhaps the most powerful continuum theory for NLCs \cite{dg,virga1995variational,lin2001static}. In 1991, Pierre--Gilles de Gennes was awarded the Nobel prize in Physics for discovering that "methods developed for studying order phenomena in simple systems can be generalized to more complex forms of matter, in particular to liquid crystals and polymers".
The LdG model describes the NLC state by a macroscopic order parameter---the LdG $\Q$-tensor, which is a macroscopic measure of NLC orientational order, i.e., the deviation of the ordered nematic phase from the isotropic disordered phase. Mathematically, the $\Q$-tensor is a symmetric traceless $3\times3$ matrix.
The $\Q$-tensor has five degrees of freedom and can be written as \cite{newtonmottram},
\begin{equation}
\Q = \lambda_1 \n\otimes\n + \lambda_2\m\otimes\m + \lambda_3\p\otimes\p,
\end{equation}
where $\n$, $\m$, and $\p$ are eigenvectors of $\Q$ which model the nematic directors, and $\lambda_i$, $i = 1,2,3$ are the corresponding eigenvalues, which measure the degree of orientational order about these directors. In particular, $\sum_{i = 1}^3\lambda_i =0$, from the tracelessness constraint. 
A $\Q$-tensor is said to be (i) isotropic if~$\Q=0$, i.e., $(\lambda_1,\lambda_2,\lambda_3) = (0,0,0)$, (ii) uniaxial if $\Q$
has a pair of degenerate non-zero eigenvalues, $(\lambda, \lambda, - 2\lambda)$, and (iii) biaxial if~$\Q$ has three distinct eigenvalues \cite{dg}.
%,i.e., $\lambda_1\neq\lambda_2\neq\lambda_3$~\cite{dg}. 
A uniaxial $\Q$-tensor can be written as
\begin{equation}\label{eq:uniaxial_Q}
\Q = s \left(\n \otimes \n - \I/3\right)
\end{equation} 
with~$\I$ being the $3\times 3$ identity matrix, $s = -3\lambda$ is  real and~$\n\in \mathbb{S}^2$, a unit vector. The vector, $\n$, is the eigenvector with the non-degenerate eigenvalue, known as the "director" and models the single preferred direction of uniaxial nematic alignment at every point in space~\cite{virga1995variational,dg}. The scalar, $s$, is the scalar order parameter, which measures the degree of orientational order about $\n$. In the biaxial case, there is a primary and a secondary nematic director, with two scalar order parameters.

In the absence of surface energies, the LdG energy is given by
\begin{equation}
    I_{LdG}[\Q]:=\int f_{el}(\Q,\nabla\Q) + f_b\left(\Q\right) \mathrm{d}\x,\label{eq:3Denergy}
\end{equation}
where $f_{el}$ and $f_b$ are the elastic and thermotropic bulk energy densities, respectively.
The elastic energy density is typically quadratic and convex in $\nabla \Q$, 
 and penalises spatial inhomogeneities. In general, the elastic energy density has different contributions from different deformation modes e.g. splay, twist and bend \cite{dg}. A commonly used version is
\begin{equation}\label{eq:fel_full}
    f_{el}(\Q) = \frac{L_1}{2}Q_{ij,k}Q_{ij,k}+ \frac{L_2}{2}Q_{ij,j}Q_{ik,k} + \frac{L_3}{2}Q_{ik,j}Q_{ij,k}
\end{equation}
where $L_1$, $L_2$, $L_3$ are material elastic constants, subject to certain constraints to ensure $f_{el}(\Q) \geq 0$. Since 
\begin{equation}
 Q_{ij,j}Q_{ik,k}-Q_{ik,j}Q_{ij,k} = (Q_{ij}Q_{ik,k}),j - (Q_{ij}Q_{ik,j}),k
\end{equation}
is a null Lagrangian, we can ignore the $L_3$-term with Dirichlet boundary conditions.
Hence, the elastic energy density in \eqref{eq:fel_full} is reduced to a two-term elastic energy density, as shown below %with $L_1 = L$ given by
\begin{equation}\label{eq:fel}
    f_{el}(\Q)=\frac{L}{2}\left(|\nabla\Q|^2+\hat{L}_2(\div{\Q})^2\right),
\end{equation}
where $\hat{L}_2\in(-1,\infty)$ is the "elastic anisotropy" parameter. The elastic anisotropy can be strong for polymeric materials \cite{wensink2019polymeric}. In Section \ref{sec:ani}, we study the effects of elastic anisotropy on NLC solution landscapes on square domains.

In Section \ref{sec:pol}, \ref{sec:rec}, \ref{sec:SL}, we use the one-constant approximation, for which, $L_2= L_3 = 0$ in (\ref{eq:fel_full}) i.e. $\hat{L}_2 = 0$ in \eqref{eq:fel}, so that the elastic energy density simply reduces to the Dirichlet energy density $|\nabla \Q|^2$. The one-constant approximation assumes that all deformation modes have comparable energetic penalties i.e. equal elastic constants and this is a good approximation for some characteristic NLC materials such as MBBA \cite{dg,virga1995variational}, which makes the mathematical analysis more tractable.

The bulk energy density $f_b$ is a polynomial of the eigenvalues of order parameter $\Q$, and drives the isotropic-nematic phase transition as a function of the temperature~\cite{newtonmottram,dg}. We work with the simplest form of $f_b$, a quartic polynomial of eigenvalues of $\Q$-tensor:
\begin{gather}
    f_b(\Q):=\frac{A}{2}\tr\Q^2-\frac{B}{3}\tr\Q^3+\frac{C}{4}(\tr\Q^2)^2. \label{bulk}
\end{gather}
where, $\tr\Q^2=Q_{ij}Q_{ij} = \lambda_i^2$, and $\tr\Q^3=Q_{ij}Q_{jk}Q_{ki} = \lambda_i^3$, for $i,j,k=1,2,3$. The variable $A = \alpha\left(T-T^*\right)$ is a rescaled temperature, $\alpha$, $L$, $B$, $C>0$ are material-dependent constants, and $T^*$ is the characteristic nematic supercooling temperature. 
The rescaled temperature $A$ has three characteristic values: (i) $A = 0$, below which the isotropic phase $\Q = 0$ loses stability, (ii) the nematic-isotropic transition temperature, $A = B^2/27C$, at which $f_b$ is minimized by the isotropic phase and a continuum of uniaxial states with $s = s_+ = B/3C$ and $\n$ arbitrary in \eqref{eq:uniaxial_Q}, and (iii) the nematic superheating temperature, $A = B^2/24C$ above which the isotropic state is the unique critical point of $f_b$.

As Proposition $1$ in \cite{majumdar2010equilibrium}, for a given $A<0$, the set of minima of the bulk potential is 
\begin{equation}
\mathscr{N}:=\{\Q \in S_0:\Q=s_+\left(\n\otimes \n-\I/3\right)\},
\end{equation}
where
\begin{equation}
    s_+:=\frac{B+\sqrt{B^2+24|A|C}}{4C}\nonumber
\end{equation}
and $\n\in S^2$ arbitrary. In particular, this set is relevant to our choice of Dirichlet conditions for boundary-value problems in subsequent sections.
The size of defect cores is typically inversely proportional to $s_+$ for low temperatures $A<0$.  Following \cite{wojtowicz1975introduction}, we use MBBA as a representative NLC material and use its reported values for $B$ and $C$ to fix $B=0.64\times10^4 N/m^2$ and $C=0.35\times10^4 N/m^2$ throughout this review. We also frequently use the fixed temperature, $A = -B^2/3C$ for numerical simulations, although the qualitative conclusions remain unchanged for $A<0$. %We fix the rescaled temperature $A = -B^2/3C$, in order to compare the results in reduced LdG theory with reduced order parameter and LdG theory. The reduced LdG theory will be illustrated in Section \ref{sec:pol}.

Boundary effects are a crucial consideration for NLCs in confinement, and dictate multistability to some extent. %important for NLC in confinement. 
There are multiple mathematical choices for the boundary conditions. The simplest approach is Dirichlet boundary conditions or fixed boundary conditions for the LdG $\Q$-tensor order parameter. This fixes the nematic directors and the scalar order parameters on the boundary. %approach is to use Dirichlet boundary condition. We explicitly prescribe the order parameter in LdG theory, i.e., the preferred molecular alignment and nematic order in this direction, on the boundary. 
Typically, we impose tangential or homeotropic boundary condition, which means the nematic director is tangent or normal to the domain boundary. On domains with sharp corners, some care is needed to deal with the mismatch in the nematic director at the corners. This could involve truncating the geometry or imposing a low-order point at the sharp corners. %We deal with the boundary condition in mismatching region carefully in order to reduce the singular behaviour near the corners. 
Dirichlet conditions are mathematically more tractable but weak anchoring is more realistic, with surface energies and the resulting boundary conditions typically involve the normal derivatives of $\Q$ on the boundary. % that impose preferential anchoring  approach. One can employ a surface anchoring energy which enforces a preferred LdG $\Q$ on the boundary. 
A popular surface energy, known as the Rapini-Papoular energy,  is \cite{virga1995variational}
\begin{equation}
E_s[\Q] = \int_{\partial} W \tr(\Q-\Q_s)^2 dA,
\end{equation}
where $W$ is the surface anchoring strength and $\Q_s$ is the preferred LdG $\Q$-tensor on the boundary. As $W\to\infty$, we qualitatively recover the Dirichlet condition $\Q=\Q_s$ on the boundary.
Interested readers are referred to \cite{luo2012multistability}.
%In this review, we only focus on tangent Dirichlet boundary conditions for 2D domains.

We model nematic profiles inside three-dimensional wells
\begin{equation}
\label{eq:domain}
\mathcal{B} = \Omega \times \left[0, h \right],
\end{equation}  
whose cross-section is a two-dimensional polygon $\Omega$, and $h$ is the well height. The two-dimensional working domain $\Omega$ is any regular polygon in Section \ref{sec:pol}, a rectangle in Section \ref{sec:rec}, a square in Section \ref{sec:ani}, and a regular hexagon in Section \ref{sec:SL}.
In the thin film limit, i.e., $h \to 0$ limit and for certain choices of the surface energies, we can rigorously justify the reduction from the three-dimensional domain $\mathcal{B}$ to the two-dimensional domain $\Omega$  \cite{Golovaty2015Dimension}. If we impose a Dirichlet boundary condition, $\Q_b$, which has the unit-vector, $\z=(0,0,1)$ as a fixed eigenvector, on the lateral surfaces, $\partial \Omega \times \left[0, h \right]$, then one can show that in the $\frac{h}{\lambda}\to 0$  limit, where $\lambda^2$ is a measure of the cross-section size, minima of the LdG energy (\ref{eq:3Denergy}) converge (weakly in~$H^1$)
 to minima of the reduced functional
% \frac{L}{2}\left(|\nabla\Qvec|^2+\hat{L}_2(\Div{\Qvec})^2\right)
\begin{equation}\label{eq:reduced}
F_0[\Q] := \int_{\Omega} \frac{1}{2}\left(\left|\nabla_{x,y}\Q\right|^2 + \hat{L}_2\left(\div_{x,y}\Q\right)^2\right) + \frac{\lambda^2}{L} f_b\left(\Q\right) \mathrm{dA}
\end{equation}
subject to the boundary condition $\Q = \Q_b$ on $\partial\Omega$ and to the constraint that $\z$ is an eigenvector of $\Q\left(x, \, y\right)$  for any $\left(x, \, y\right)\in\Omega$.
 %subject to the boundary condition converge (weakly in~$H^1$)
 %to minima of the reduced functional
% \frac{L}{2}\left(|\nabla\Qvec|^2+\hat{L}_2(\Div{\Qvec})^2\right)
%\begin{equation}\label{eq:reduced}
Using the reasoning above, we restrict ourselves to $\Q$-tensors with $\z$ as a fixed eigenvector  
 and study critical points or minima of (\ref{eq:reduced}) with three degrees of freedom as -
\begin{equation}
    \begin{aligned}
        \Q\left(x,y\right) &= q_1\left(x,y\right)\left(\xhat\otimes\xhat-\yhat\otimes\yhat\right) + q_2\left(x,y\right)\left(\xhat\otimes\yhat+\yhat\otimes\xhat\right)\\
        &+ q_3\left(x,y\right)\left(2\zhat\otimes\zhat-\xhat\otimes\xhat-\yhat\otimes\yhat\right)
    \label{q123}
    \end{aligned}
\end{equation}
where $\xhat$, $\yhat$ and $\zhat$ are the unit coordinate vectors in the $x$, $y$ and $z$ directions respectively.
Informally speaking, $q_1$ and $q_2$ measure the degree of "in-plane" order, $q_3$  measures the "out-of-plane" order and $\Q$ is invariant in the $z$-direction.
This constraint naturally excludes certain solutions such as the stable escaped radial with ring defect solution in a cylinder with large radius in \cite{han2019transition}, for which the $z$-invariance does not hold or critical points that exhibit "escape into the third dimension" \cite{sonnet1995alignment}, for which $\zhat$ is not a fixed eigenvector for $\Q$. Whilst we present our results in a 2D framework in the case studies, these reduced critical points survive for all $h>0$ (beyond the thin-film limit) although they may not be physically relevant or energy-minimizing outside the thin-film limit (\cite{canevari2017order} and \cite{wang2019order}).% \textcolor{red}{the stable escape radial with ring defect solution and the stable escaped radial solution in a cylinder with large radius in \cite{han2019transition}}, for which the $z$-invariance does not hold or critical points that exhibit "escape into the third dimension" \cite{sonnet1995alignment}, for which $\zhat$ is not a fixed eigenvector for $\Q$. Whilst we present our results in a 2D framework in the case studies, these reduced critical points survive for all $h>0$ (beyond the thin-film limit) although they may not be physically relevant or energy-minimizing outside the thin-film limit (\cite{canevari2017order} and \cite{wang2019order}). 

\section{Benchmark Example}\label{sec:ben}
The square domain is a very well-studied domain and we review some classical results in this section. %. In this review, we use the results of nematic liquid crystal confined on a square in the LdG framework without elastic anisotropy as a benchmark.
In \cite{kralj2014order} and \cite{canevari2017order}, the authors report the Well Order Reconstruction Solution ($WORS$) on a square domain, for all square edge lengths $\lambda >0$, without elastic anisotropy, for Dirichlet tangent boundary conditions. 
The $WORS$ has a constant set of eigenvectors, $\xhat$, $\yhat$, and $\zhat$, which are the coordinate unit vectors. The $WORS$ is further distinguished by a uniaxial cross, with negative scalar order parameter, along the square diagonals. Physically, this implies that there is a planar defect cross along the square diagonals, and the nematic molecules are disordered along the square diagonals. This defect cross partitions the square domain into four quadrants, and the nematic director is constant in each quadrant. The defect cross is an interesting example of a negatively ordered uniaxial interface that separates distinct polydomains.  In \cite{canevari2017order}, the authors analyse this system at a fixed temperature $A=-B^2/3C$, and show that the $WORS$ is a classical solution of the associated Euler--Lagrange (EL) equations for the LdG free energy, of the form:
\begin{align}
\mathbf{Q}_{WORS}(x,y)=q(\xhat\otimes\xhat-\yhat\otimes\yhat)-\frac{B}{6C}(2\zhat\otimes\zhat-\xhat\otimes\xhat-\yhat\otimes\yhat). \label{WQ}
\end{align}
There is a single degree of freedom, $q:\Omega\to\mathbb{R}$, which satisfies the Allen-Cahn equation 
\begin{equation}
    \Delta q = \frac{\lambda^2}{L}(2Cq^3-\frac{B^2}{2C}q)
\end{equation}
and exhibits the following symmetry properties:
\begin{gather}
    q=0\quad\text{on}\quad\{y=x\}\cup\{y=-x\},\qquad
    (y^2-x^2)q(x,y)\geq0.
\end{gather}
 Mathematically speaking, this implies that the $\Q_{WORS}$ is strictly uniaxial with negative order parameter along the square diagonals which would manifest as a pair of orthogonal defect lines in experiments. 
They also prove that the $WORS$ is globally stable for $\lambda$ small enough i.e. nano-scale domains, and becomes unstable as $\lambda$ increases, demonstrating a pitchfork bifurcation in a scalar setting. Numerical experiments suggest that the $WORS$ acts as a transition state between energy minimizers for large $\lambda$. For large square domains (on micron scale or larger), there are two competing stable physically observable states: the largely uniaxial diagonal  states ($D$), for which the nematic director (in the plane) is aligned along one of the square diagonals and the rotated states ($R$) for which the director rotates by $\pi$ radians between a pair of opposite square edges. On a  square domain, there are $2$ rotationally equivalent $D$ states, and $4$ rotationally equivalent $R$ states \cite{kusumaatmaja2015free, lewis2014colloidal}. We note that the $D$ and $R$ states have non-zero $q_2$ in (\ref{q123}) whilst the $WORS$ has $q_2=0$ everywhere. In other words, the $WORS$ solution has constant eigenvectors everywhere whereas the $D$ and $R$ solutions have varying eigenvectors in the plane of the square domain.

\begin{figure}[b]
    
    \includegraphics[width=0.649\columnwidth]{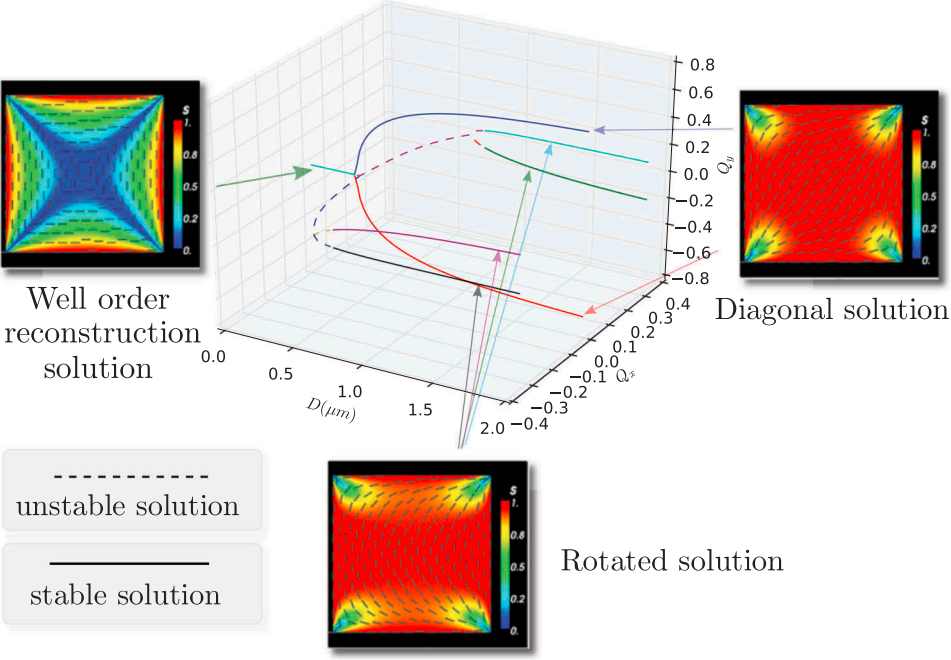}
    \caption{The bifurcation diagram on square without elastic anisotropy. Reproduced from \cite{robinson2017molecular} with permission from Taylor$\&$Francis}
    \label{benchmark}
\end{figure}

The bifurcation diagram for this model problem, has been documented in \cite{robinson2017molecular} (see Fig. \ref{benchmark}). For $\lambda < \lambda^*$, there is the unique $WORS$. For $\lambda = \lambda^*$, the stable $WORS$ bifurcates into an unstable $WORS$, and two stable $D$ solutions. When $\lambda = \lambda^{**}>\lambda^*$, the unstable $WORS$ bifurcates into two unstable $BD$ solutions, which are featured by defect lines localised near a pair of opposite square edges. The two $BD$ solution branches are represented by the dashed lines in Fig. \ref{benchmark}. Each unstable $BD$ solution further bifurcates into two unstable $R$ solutions, which gain stability as $\lambda$ further increases. The $WORS$ has the highest energy amongst the numerically computed solutions, for all $\lambda$.

\section{Nematic Equilibria on 2D Polygons}\label{sec:pol}
This section reviews results from a recent paper \cite{han2020pol}, where the authors study multistability for NLCs in regular 2D polygons, with tangent boundary conditions, with emphasis on the effects of geometry captured by the polygon edge length, $\lambda$. % on the effect of geometry on the nematic equilibria in two asymptotic limits -- the $\lambda\to 0$ limit of vanishing cross-section size, and $\lambda\to\infty$ limit relevant for large micron-scale systems.

\begin{figure}[b]
         \includegraphics[width=0.25\columnwidth]{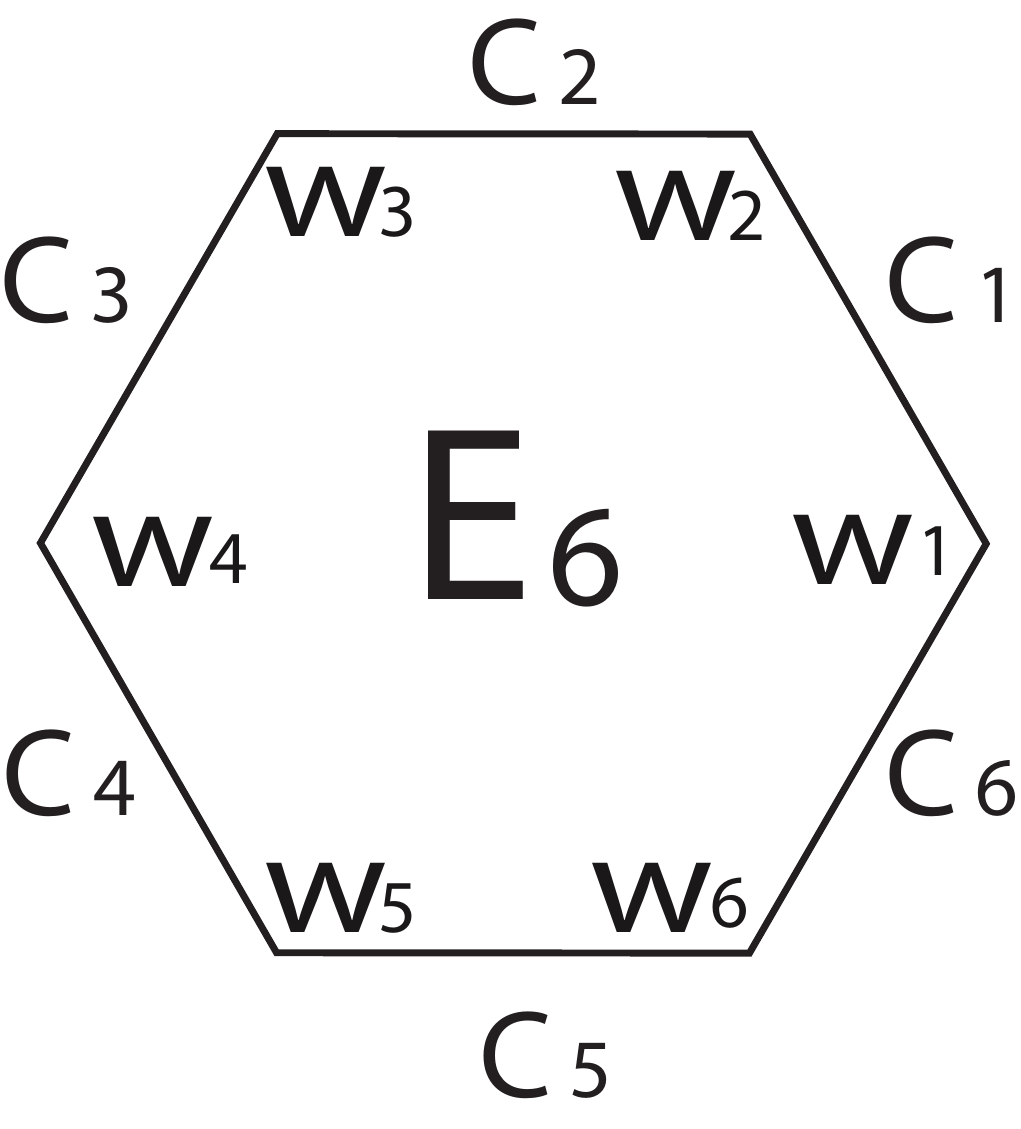}
        \caption{The regular rescaled hexagon domain $E_6$. Reproduced from \cite{han2020pol} with permission from Society for Industrial and Applied Mathematics}
        \label{domain}   
\end{figure}

The working domain, $\Omega$, is a regular rescaled polygon, $E_K$, with $K$ edges, centered at the origin with vertices
\begin{equation}
    w_k =  \left(\cos\left(2\pi \left(k-1\right)/K\right),\sin\left(2\pi \left(k-1\right)/K\right)\right),\ k = 1,...,K.\nonumber
\end{equation}
We label the edges counterclockwise as $C_1, ..., C_K$, starting from $\left(1,0\right)$. For example $E_6$ is a regular hexagon shown in Fig. \ref{domain} and $E_4$ is a square.

As elaborated in Section~\ref{sec:theory}, it is reasonable to work with in a reduced LdG framework, with $\Q$-tensors of the form in (\ref{q123}), on 2D polygons. % work in a reduced LdG framework. 
From \cite{canevari_majumdar_wang_harris}, for the special temperature $A = -B^2/3C$, we necessarily have $q_3 = -\frac{B}{6C}$, for all $\lambda>0$. For arbitrary $A<0$, we would have non-constant $q_3$ profiles and whilst we conjecture that some qualitative solution properties are universal for $A<0$, a non-constant $q_3$ profile would introduce new technical difficulties. 
For $A = -B^2/3C$ and with constant $q_3$, the $\Q$-tensor in (\ref{q123}) reduces to a symmetric, traceless $2\times 2$ matrix, $\P$, as given below -
\begin{equation}
    \P =
    \left(\begin{tabular}{cc}
    $P_{11}$&$P_{12}$\\
    $P_{12}$&$-P_{11}$\\
\end{tabular}\right).
\nonumber
\end{equation}
The relation between the LdG order parameter $\Q$-tensor and the reduced $\P$-tensor is
\begin{equation}
    \Q =
    \left(\begin{tabular}{cc|c}
        \multicolumn{2}{c|}{\multirow{2}*{$\P\left(\r\right)+\frac{B}{6C}\I_2$}} & $0$ \\
    \multicolumn{2}{c|}{} & $0$ \\
    \hline
    $0$ & $0$ & $-B/3C$ \\
    \end{tabular}\right).
    \label{eq:QP}
\end{equation}
Therefore, the energy in (\ref{eq:reduced}) is reduced to
    \begin{equation}
        F[P]: = \int_{\Omega}\frac{1}{2}|\nabla P|^2+\frac{\lambda^2}{L}\left(-\frac{B^2}{4C}tr\P^2+\frac{C}{4}\left(tr\P^2\right)^2\right) \mathrm{d A},
        \label{p_energy}
    \end{equation}
and the corresponding EL equations are
\begin{equation}
    \begin{aligned}
        \Delta P_{11} &= \frac{2C\lambda^2}{L}\left(P_{11}^2+P_{12}^2-\frac{B^2}{4C^2}\right)P_{11},\\
        \Delta P_{12} &= \frac{2C\lambda^2}{L}\left(P_{11}^2+P_{12}^2-\frac{B^2}{4C^2}\right)P_{12}.\\
\end{aligned}
    \label{Euler_Lagrange}
\end{equation}
We can also write $\P$ in terms of an order parameter $s$ and an angle $\gamma$ as shown below -
\begin{equation}
    \P = 2s\left(\n\otimes\n-\frac{1}{2}\I_2\right)\label{P}; \quad \n = \left(\cos\gamma,\sin\gamma\right)^T,
\end{equation}
where 
%\begin{equation}\label{eq:n}
%\n = \left(\cos\gamma,\sin\gamma\right)^T
%\end{equation}
$I_2$ is the $2\times 2$ identity matrix, 
so that
$ P_{11} = s\cos\left(2\gamma\right),\ P_{12} = s\sin\left(2\gamma\right).$
The nodal set, defined by the zeroes of $\P$, model the planar defects in $\Omega$ i.e. when $\P=0$, $s=0$ in (\ref{P}) so that there is no nematic order in the plane of $\Omega$, and the eigenvalues of the corresponding $\Q$ are $(B/6C,B/6C,-B/3C)$. In other words, the nodal set of $\P$ defines a uniaxial set of $\Q$ with negative order parameter, and will have a distinct optical signature in experiments. %We can accurately capture the structure and location of the optical defect, which is mathematically identified with the zero set of the reduced solution. The zero simply means there is no nematic order in the plane of $\Omega$.

Next, we specify Dirichlet tangent boundary conditions for $\P$ on $\partial E_K$, labelled by $\P_b$. The tangent boundary conditions require $\n$ in (\ref{P}) to be tangent to the edges of $E_K$, and $s = s_+ = B/3C$. However, there is a necessary mismatch at the vertices, so that we fix the value of $\P$ at the vertex, to be the average of the two constant values, on the two intersecting edges. On a $d \ll \frac{1}{2}$-neighbourhood of the vertices, we linearly interpolate between the constant values on the edge and the average value at the vertex. For $d$ sufficiently small, the choice of the interpolation does not change the qualitative solution profiles. This means that the tangent conditions are not necessarily respected in the $d$-neighbourhood of vertices.

In what follows, we study the minima of (\ref{p_energy}) in two distinguished limits analytically -- the $\lambda \to 0$ limit is relevant for nano-scale domains and the $\lambda\to \infty$ limit, which is the macroscopic limit relevant for micron-scale or larger cross-sections, $\Omega$. 
We present rigorous results for limiting problems below but our numerical simulations show that the limiting results are valid for non-zero but sufficiently small $\lambda$ (or even experimentally accessible nano-scale geometries depending on parameter values) and sufficiently large but finite $\lambda$ too. In other words, these limiting results are of potential practical value too.

\begin{figure}
    \centering
        \includegraphics[width=0.8\columnwidth]{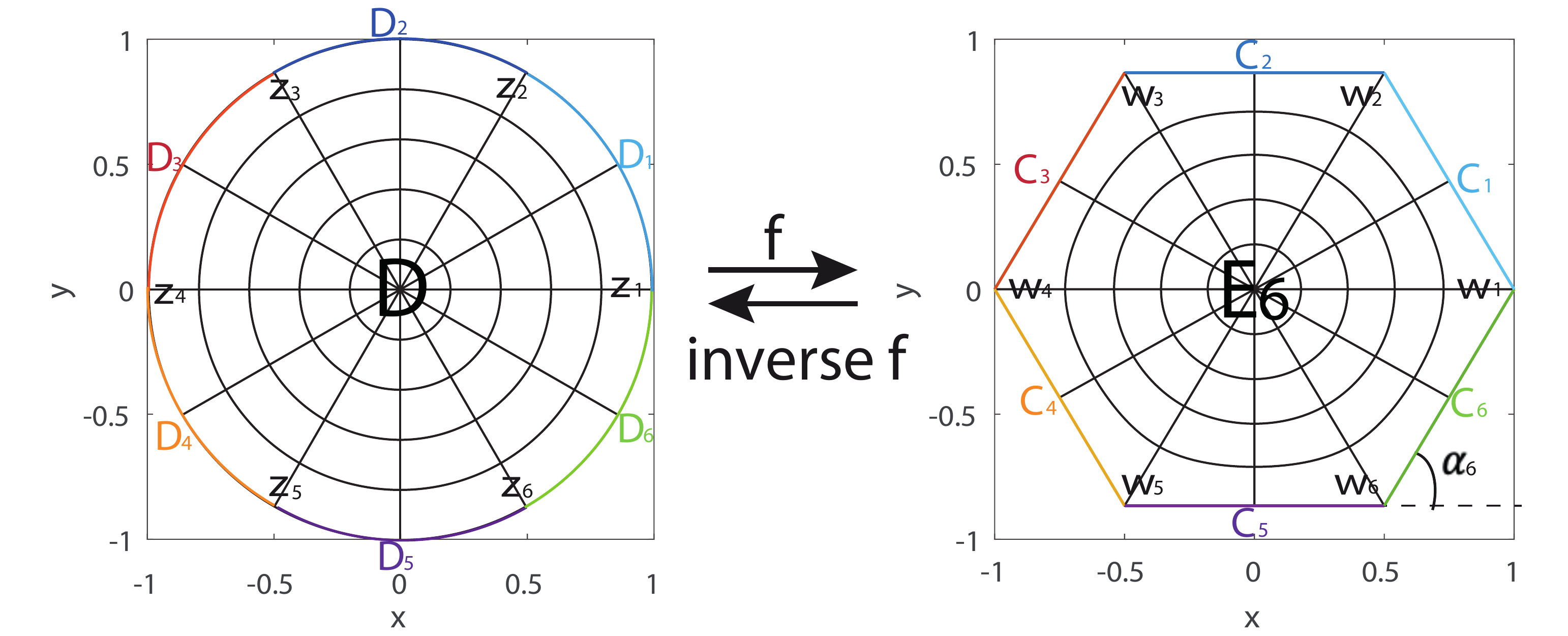}
        \caption{Schwarz--Christoffel mapping $f$ from a unit disc to a regular hexagon and inverse mapping $f^{-1}$ from a regular hexagon to a unit disc. Reproduced from \cite{han2020pol} with permission from Society for Industrial and Applied Mathematics}
        \label{circle_6_hexagon}
\end{figure}

In the $\lambda\to 0$ limit, using methods from \cite{bethuel1994ginzburg} and from Proposition $3.1$ of \cite{fang2019solution}, we can show that minima of (\ref{p_energy}), subject to the Dirichlet tangent boundary conditions (for $d$ sufficiently small), converge uniformly to the unique solution of the following limiting problem for $\lambda = 0$,
\begin{equation}
\begin{aligned}
    &\Delta P_{11}^0 = 0,\ \Delta P_{12}^0 = 0, on\ E_K,\\
    &P_{11}^0 = P_{11b},\ P_{12}^0 = P_{12b},\ on\ \partial E_K.
\end{aligned}
    \label{zero_euler}
\end{equation}
The solution of Laplace equation on disc can be explicitly solved. Our strategy to solve the Dirichlet boundary-value problem \eqref{zero_euler} on polygon $E_K$ is to map it to an associated Dirichlet boundary-value problem on the unit disc in Fig. \ref{circle_6_hexagon}, by using the Schwarz--Christoffel mapping~\cite{brilleslyper2012explorations}.
The SC mapping from a unit disc to a regular polygon $E_K$ is
\begin{equation}
    f\left(z\right) = C_1\left(K\right)\int_0^z\frac{1}{\left(1-x^K\right)^{2/K}}\rm{d}x\nonumber
\end{equation} with
\begin{equation}
    C_1\left(K\right) = \frac{\Gamma\left(1-1/K\right)}{\Gamma\left(1+1/K\right)\Gamma\left(1-2/K\right)}.\nonumber
\end{equation}

\begin{figure}
    \begin{center}
        \includegraphics[width=\columnwidth]{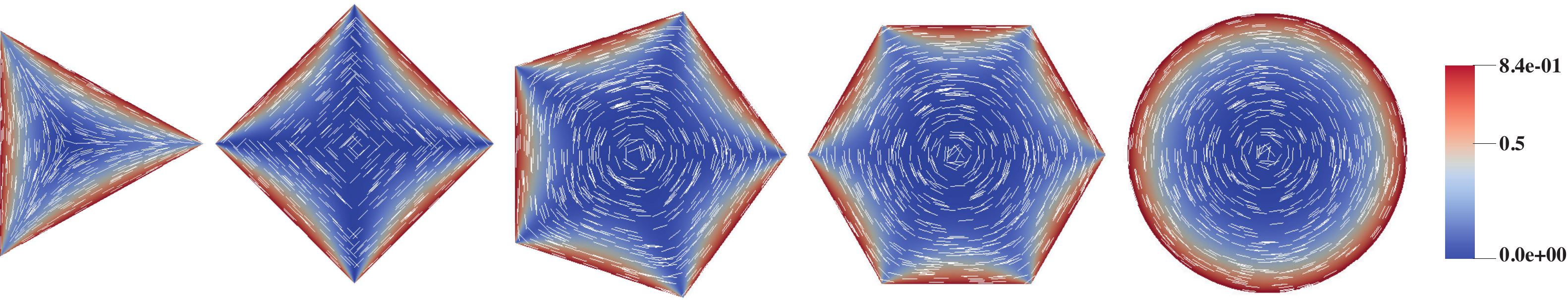}
        \caption{Solutions $\left(P_{11}^0,P_{12}^0\right)$ of (\ref{zero_euler}) when $K = 3,4,5,6$ in regular triangle, square, pentagon, hexagon domain and $K\to\infty$ in disc domain.
        The vector $\left(\cos\left(\arctan\left(P_{12}^0/P_{11}^0\right)/2\right),\sin\left(\arctan\left(P_{12}^0/P_{11}^0\right)/2\right)\right)$ is represented by white lines and the order parameter $\left(s^0\right)^2=\left(P_{11}^0\right)^2 + \left(P_{12}^0\right)^2$ is represented by color from blue to red. Reproduced from \cite{han2020pol} with permission from Society for Industrial and Applied Mathematics}
        \label{direction_multi_edge}
    \end{center}
\end{figure}

Using the symmetries of the boundary condition and the regular polygon, we can prove the symmetry properties of the limiting solution of \eqref{zero_euler} accompanied by rigorous results for the corresponding nodal set, as given below (from \cite{han2020pol}).
\begin{proposition} Let $\left(P_{11}, P_{12} \right)$ be the unique solution of (\ref{zero_euler}) and let
    \begin{equation}
        G_K:=\{S\in O\left(2\right):SE_K\in E_K\},
        \label{sym-set}
    \end{equation}
be a set of symmetries consisting of
$K$ rotations by angles $2\pi k/K$ for $k = 1,...,K$ and $K$ reflections about the symmetry axes ($\phi=\pi k/K$, $k = 1,...,K$) of the polygon $E_K$.\\
$P_{11}^2+P_{12}^2$ is invariant under $G_K$.
    If $\left(P_{11},P_{12}\right)\neq (0,0)$, then
    $\frac{\left(P_{11},P_{12}\right)}{\sqrt{P_{11}^2+P_{12}^2}}$ undergoes a reflection about the symmetry axes of the polygon and rotates by $4\pi k/K$ under rotations of angle $2\pi k/K$ for $k = 1,...,K$.
    \label{symmetry_proposition}
\end{proposition}
\begin{proposition} Let $\P_R= \left(P_{11}, P_{12}\right)$ be the unique solution of the boundary-value problem (\ref{zero_euler}). Then $P_{11}\left(0,0\right) = 0, P_{12}\left(0,0\right) = 0$ at the centre of all regular polygons, $E_K$.
However, $\P_R \left(x, y\right) \neq \left(0,0\right)$ for $\left(x,y\right)\neq \left(0,0\right)$, for all $E_K$ with $K \neq 4$ i.e. the $WORS$ is a special case of $\P_R$ on $E_4$ such that $\P_R = \left(0,0\right)$ on the square diagonals. 
%For $K\neq 4$, the origin is the unique zero of the unique solution $\P_R$, referred to as the "Ring solution" in the rest of the section.
    \label{isotropic_center}
\end{proposition}
%The square, $E_4$ is special since the eigenvectors of the associated $\P_R$ are constant in space and $\P_R$ vanishes along the square diagonals. 
For $K\neq 4$, $\P_R$ has a unique isotropic point at the origin and is referred to as the $Ring$ solution, since for $K>4$, the director profile (the profile of the leading eigenvector of $\P_R$ with the largest positive eigenvalue) exhibits a $+1$-vortex at the centre of the polygon. In Fig. \ref{direction_multi_edge}, we numerically plot the ring configuration for a triangle, pentagon, hexagon and a disc ($K\to\infty$) and the $WORS$ for the square.
For $K=3$, the isotropic point at the centre of the equilateral triangle resembles a $-1/2$ point defect. This is a very interesting example of the effect of geometry on solutions, and their defect sets. % with profound optical and experimental implications.

The $\lambda\to\infty$ limit is analogous to the "Oseen--Frank limit" in \cite{majumdar2010landau}. Let $\P^{\lambda}$ be a global minimizer of (\ref{p_energy}), subject to a fixed boundary condition $\P_b = \left(P_{11b},P_{12b}\right)$ on $\partial E_K$. As $\lambda \to \infty$, the minima, $\P^{\lambda}$, converge strongly in $W^{1,2}$ to $\P^{\infty}$ where
    \begin{equation}
        \P^{\infty} = \frac{B}{2C}\left(\n^{\infty}\otimes \n^{\infty}-\frac{1}{2}\I_2\right),\nonumber
    \end{equation}
    $\n^{\infty} = \left(\cos\gamma^{\infty},\sin\gamma^{\infty}\right)$ and $\gamma^{\infty}$ is a global minimizer of the energy
    \[
        I[\gamma]: = \int_{E_K} \left| \nabla \gamma \right|^2 \mathrm{dA}
    \]
subject to Dirichlet conditions, $\gamma=\gamma_b$ on $\partial E_K$. The angle $\gamma_b$ is determined by the fixed boundary condition, $\P_b$, where  $\n_b = \left(\cos\gamma_b, \sin\gamma_b \right)$. We have $\n_b$ is tangent to the polygon edges, which constrains the values of $\gamma_b$, and if $\textrm{deg}\left(\n_b,\partial E_K\right) = 0$, then $\gamma^{\infty}$ is a solution of the Laplace equation
    \begin{equation}
        \begin{aligned}
            \Delta\gamma^{\infty} &= 0,\ on\ E_K\\
        \end{aligned}
        \label{infty_euler}
    \end{equation}
subject to $\gamma=\gamma_b$ on $\partial E_K$ \cite{lewis2014colloidal,bethuel1993asymptotics}.

\begin{figure}
    
        \includegraphics[width=0.4\columnwidth]{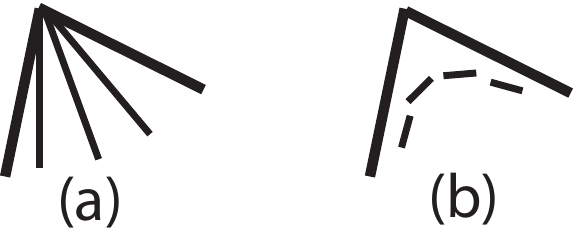}
        \caption{Two arrangements of nematics in the corner: (a) splay and (b) bend. Reproduced from \cite{han2020pol} with permission from Society for Industrial and Applied Mathematics}
        \label{normal_angle}
\end{figure}

There are multiple choices of $\gamma_b$ consistent with the tangent boundary conditions, which implies that there are multiple local/global minima of (\ref{p_energy}) for large $\lambda$. We present a simple estimate of the number of stable states if we restrict $\gamma_b$, so that $\n_b$ rotates by either $2\pi/K-\pi$ or $2\pi/K$ at a vertex (see Fig. \ref{normal_angle}(a) and (b), referred to as "splay" and "bend" vertices respectively). Since we require $\textrm{deg}\left(\n_b,\partial E_K\right) = 0$, we necessarily have $2$ "splay" vertices and $\left( K - 2 \right)$ "bend" vertices. 
 So we have at least $K\choose 2$ minima of (\ref{p_energy}), for $\lambda$ sufficiently large.

\begin{figure}
	\centering
    \includegraphics[width=\columnwidth]{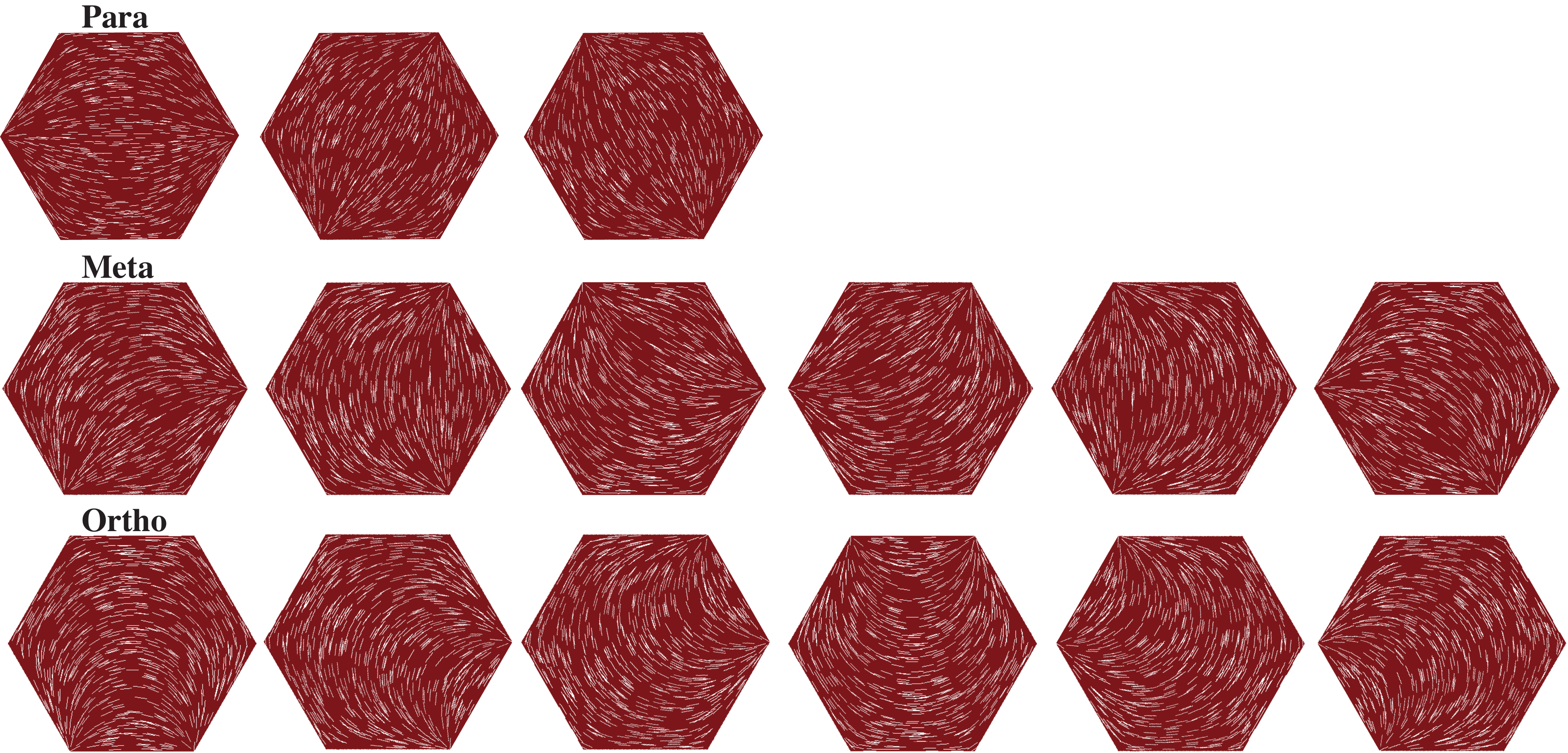}
    \caption{$6\choose 2$$=15$ solutions of (\ref{infty_euler}) subject to boundary condition (\ref{infty_boundary}) in hexagon domain.
    The vector $\left(\cos\gamma^{\infty},\sin\gamma^{\infty}\right)$ is represented by white lines. Reproduced from \cite{han2020pol} with permission from Society for Industrial and Applied Mathematics}
    \label{C62_15_states}
\end{figure}

As an illustrative example, we take the hexagon $E_6$  in Fig. \ref{C62_15_states}.
The Dirichlet boundary conditions are
\begin{equation}
    \gamma_b = \gamma_k\ on\ C_k,\ k = 1,...,K,
    \label{infty_boundary}
\end{equation}
where
\begin{equation}
    \gamma_1 = \frac{\pi}{K}-\frac{\pi}{2},\ \gamma_{k+1} = \gamma_k+jump_k,\ k = 1,2,..,K-1.\nonumber
\end{equation}
We need to choose the two splay vertices where $\gamma$ rotates as in Fig. \ref{normal_angle}(a). If the chosen corner is between the edges $C_k$ and $C_{k+1}$, then
$jump_k = 2\pi/K-\pi$, otherwise $jump_k = 2\pi/K$, $k = 1,...,K-1$. We have $15$ different choices for the two "splay" vertices, (i) $3$ of which correspond to the three pairs of diagonally opposite vertices, (ii) $6$ of which correspond to pairs of vertices which are separated by one vertex and (iii) $6$ of which correspond to "adjacent" vertices connected by an edge (see Fig. \ref{C62_15_states}). We refer to (i) as \emph{Para} states, (ii) as \emph{Meta} states and (iii) as \emph{Ortho} states.

\begin{figure}[b]
\centering
    \begin{subfigure}{0.49\textwidth}
        \centering
        \includegraphics[width=\columnwidth]{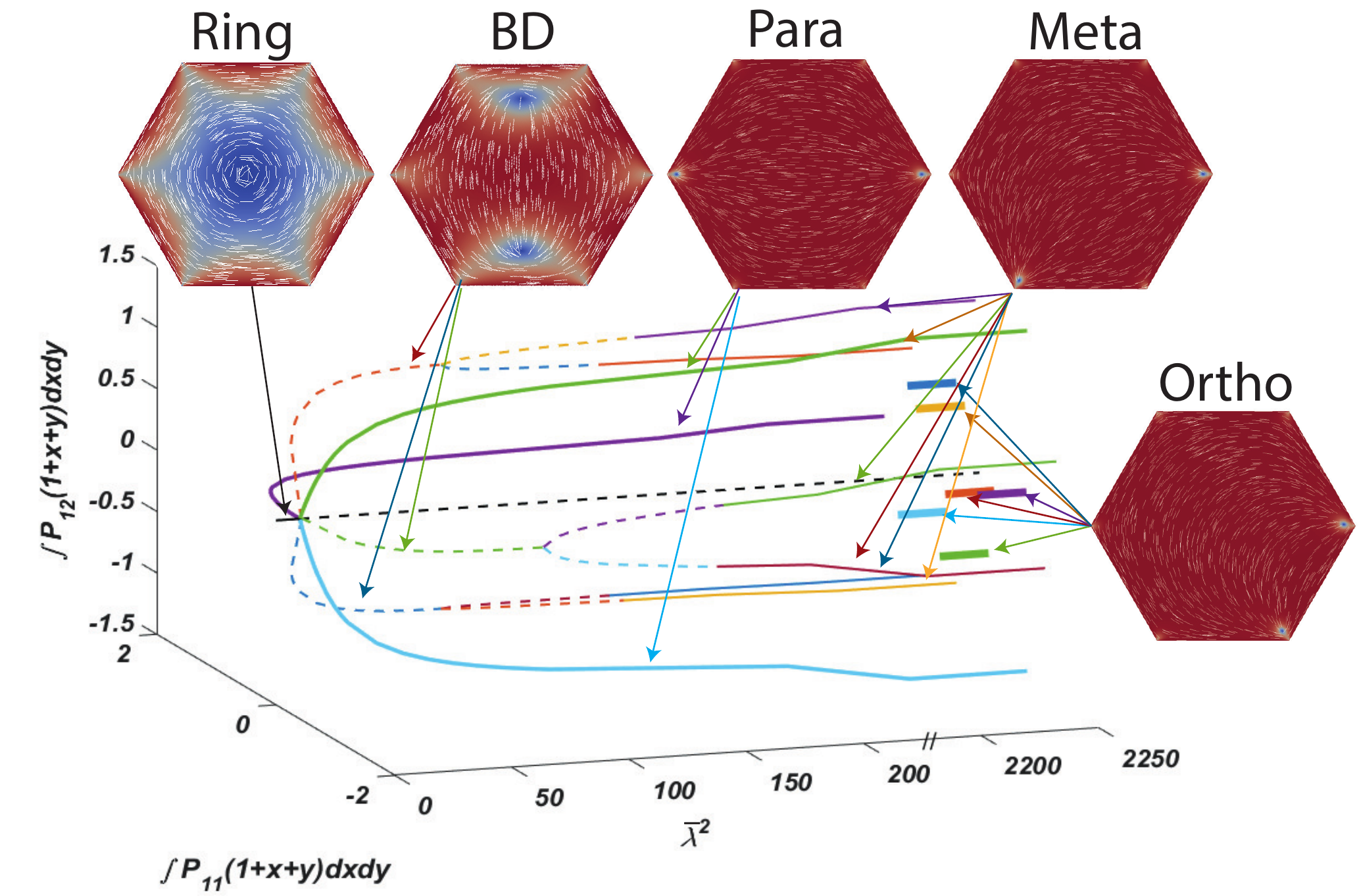}
    \end{subfigure}
        \vspace{2em}
    \begin{subfigure}{0.49\textwidth}
        \centering
        \includegraphics[width=\columnwidth]{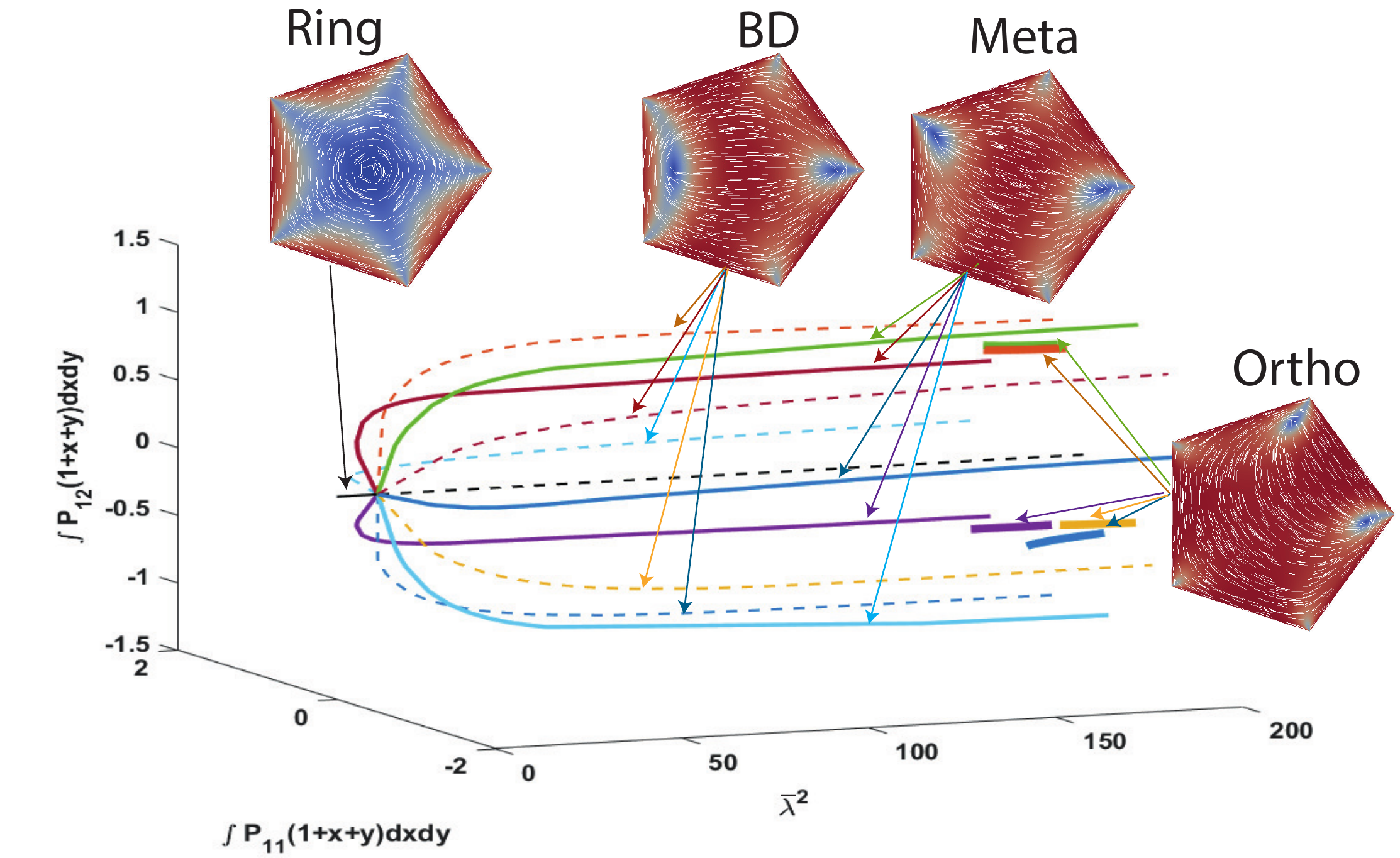}
    \end{subfigure}
    \caption{Bifurcation diagram for reduced LdG model in regular hexagon (left) and pentagon (right) domains, as a function of  $\bar{\lambda}^2= \frac{\lambda^2}{2C}$. Reproduced from \cite{han2020pol} with permission from Society for Industrial and Applied Mathematics}
    \label{hexagon_pentagon_bifurcation_diagram}
\end{figure}

Next, we present two bifurcation diagrams on a hexagon and pentagon as a function of $\lambda$, as illustrative examples of a polygon with even or odd number of edges. 
We discuss the bifurcation diagram on $E_6$ in Fig. \ref{hexagon_pentagon_bifurcation_diagram}(a). For $\lambda$ sufficiently small, there is a unique $Ring$-like minimizer. Our numerics show that the $Ring$-like solution (with the unique zero at the polygon center) exists for all $\lambda$, but there is a critical point $\lambda = \lambda^{*}$,  such that the $Ring$-like solution is unstable for $\lambda>\lambda^{*}$ and bifurcates into two kind of branches: stable $Para$ solution branches; unstable $BD$ branches. In the $BD$ state, the hexagon is separated into three regions by two "defective low-order lines" (low $|\P|^2$) such that the corresponding director (eigenvector with largest positive eigenvalue) is approximately constant in each region. There are at least three different $BD$ states. The unstable $BD$ branches further bifurcate into unstable $Meta$ solutions at $\lambda=\lambda^{**}$. There is a further critical point $\lambda = \lambda^{***}$ at which the $Meta$ solutions gain stability and continue as stable solution branches as $\lambda$ increases. Stable $Ortho$ solutions appear as solution branches for $\lambda$ is large enough. For large $\lambda$, there are multiple stable solutions: three $Para$, six $Meta$ and six $Ortho$, in Fig. \ref{C62_15_states}. The $Para$ states have the lowest energy and the $Ortho$ states are energetically the most expensive, as can be explained on the heuristic grounds that bending between neighbouring vertices is energetically unfavourable. 
The case of a pentagon is different. In Fig.~\ref{hexagon_pentagon_bifurcation_diagram}(b), there is no analogue of the $Para$ states and there are $10$ different stable states for large $\lambda$ - (i) five $Meta$ states featured by a pair of splay vertices that are separated by a vertex and (ii) five $Ortho$ states featured by a pair of adjacent splay vertices. There are five analogues of the $BD$ states which are featured by a single line of "low" order along an edge and an opposite splay vertex. 

These examples and the numerical results are not exhaustive but they do showcase the beautiful complexity and ordering transitions feasible in two-dimensional polygonal frameworks. Similar methodologies can also be applied to other non-regular polygons, convex or concave polygons. 
\section{Effects of Geometrical Anisotropy}\label{sec:rec}
The prototype problem of nematics inside square and other regular polygon domains have been discussed in Sections~\ref{sec:ben} and \ref{sec:pol}.
A natural question is what will happen if we break the symmetry of geometry? For example, what are the solution landscapes for NLCs on two-dimensional rectangles, as opposed to squares, and how sensitive is the landscape to the geometrical anisotropy? Is there the counterpart of $WORS$ on a rectangle? 

We review results from \cite{fang2020surface}. The working domain is $\Omega = [0,a]\times[0,1]$, with $a>1$, and let $\epsilon$ be a dimensionless parameter that is inversely proportional to $\lambda^2$ in the reduced LdG free energy \eqref{p_energy} in Section \ref{sec:pol}. 
We use a combination of formal calculations and elegant
maximum principle arguments to analyse solution landscapes in the  $\epsilon\to\infty$ and $\epsilon\to0$ limits. %, where $\epsilon$ is inversely proportional to $\lambda^2$ in the reduced LdG free energy \eqref{p_energy} in Section \ref{sec:pol}.

In the $\epsilon\to\infty$ limit ,i.e. the $\lambda\to0$ limit, the limiting problem is a system of Laplace equations with Dirichlet tangent boundary conditions. Analogous to the calculations in \cite{lewis2014colloidal}, the unique solution $\mathbf{P}^0$ can be calculated explicitly as $P_{12}^0 = 0$ and 
\begin{align}
P_{11}^0(x,y) &= \sum_{k\ odd}\frac{4\sin(k\pi d/a)}{k^2\pi^2 d/a}\sin\left(\frac{k\pi x}{a}\right)\frac{\sinh (k\pi(1-y)/a)+\sinh(k\pi y/a)}{\sinh(k\pi /a)} \nonumber\\
&- \sum_{k\ odd} \frac{4\sin(k\pi d)}{k^2\pi^2 d}\sin\left(k\pi y\right)\frac{\sinh (k\pi(a-x))+\sinh(k\pi x)}{\sinh(k\pi a)},
\end{align}
where $d$ is the size of mismatch region near the rectangular vertices, we linearly interpolate between the boundary conditions on the two intersecting edges to define the boundary value at the vertices.

\begin{figure}[b]
    
        \includegraphics[width=0.6\columnwidth]{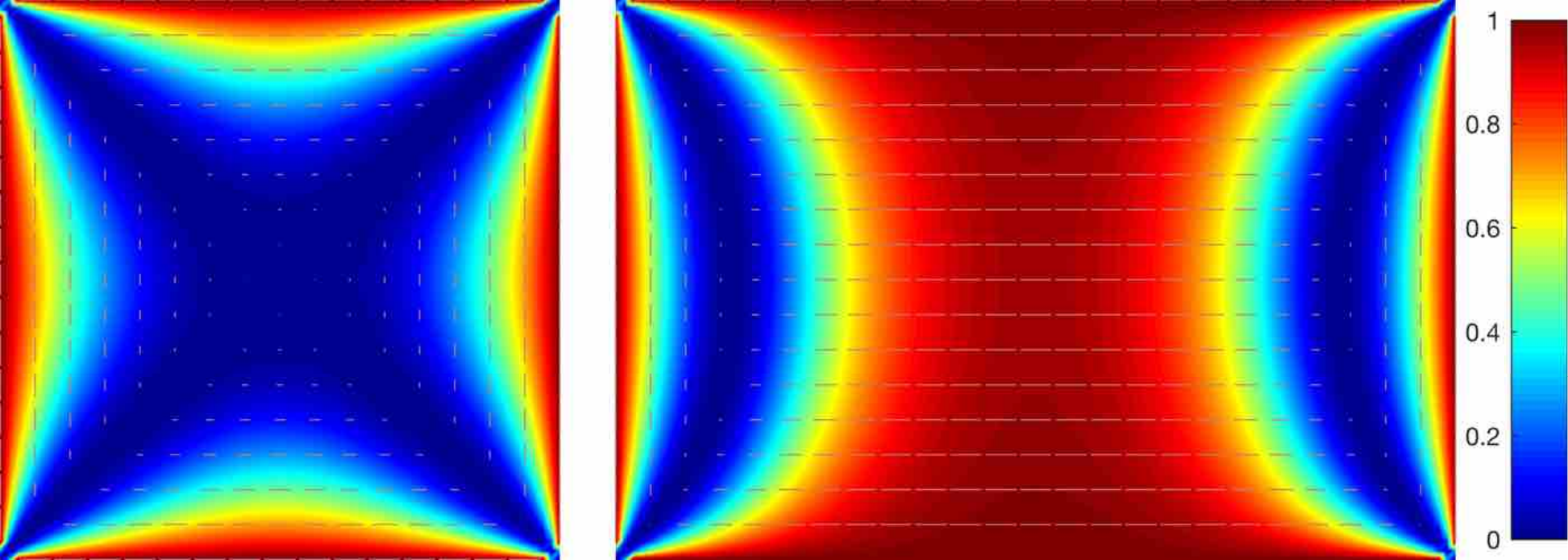}
        \caption{(Credit \cite{fang2019solution}) Left: $WORS$ on square. Right: $BD2$ on rectangle with $a = 1.5$. The color bar represents the value of $s^2 = |\mathbf{P}|/2$ in this and next figure. Reproduced from \cite{fang2019solution} with permission from SAGE Publications}
        \label{WORS_BD2}
\end{figure}

On a square with $a = 1$, the $WORS$ solution is distinguished by $\P = 0$ on the square diagonals. In fact, we can use the symmetry of the Laplace equations, Dirichlet boundary condition and the geometry, to show that $\mathbf{P}^0(1/2,1/2) = 0$. However, by constructing multiple auxiliary boundary value problem on $[0,a]\times[0,a]$, $[0,a]\times[0,1]$, $[0,1]\times[0,1]$ with suitable boundary conditions and using the maximum principle multiple times, one can prove that $P_{11}^0(a/2,1/2)> 0$ on a rectangle with $a>1$. 
The details are omitted here for conciseness, readers are referred to the elegant arguments in Proposition $3.3$ in \cite{fang2019solution}.
In the $d\to 0$ limit, the result $P_{11}^0(a/2,1/2)>0$ still holds. Hence, we lose the $WORS$ cross structure for $a\neq 1$ i.e. as soon as we break the symmetry of the square domain. %So that we lose the cross structure of the $WORS$ on all rectangles. 
In Fig. \ref{WORS_BD2}, we show the differences between the $WORS$ on a square and the limiting profile on a rectangle, labelled as $BD2$. The $BD2$ is featured by disentangled line defects near opposite short edges. 

\begin{figure}[b]
    
        \includegraphics[width=0.649\columnwidth]{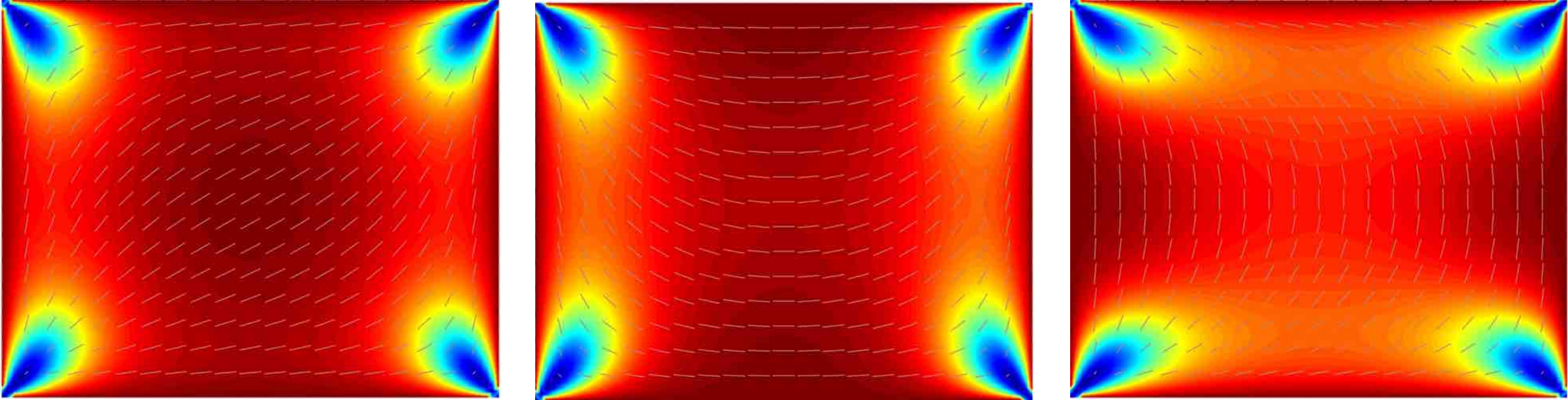}
        \caption{From left to right: $D$, $R1$, $R3$ on rectangle with $a = 1.25$. Reproduced from \cite{fang2019solution} with permission from SAGE Publications}
        \label{rectangle_D_R}
\end{figure}

In $\epsilon\to 0$ limit relevant for macroscopic domains or large $\lambda$, analogous to the approach in \cite{lewis2014colloidal}, the energy minimizers of (\ref{p_energy}) can be studied in terms of Dirichlet boundary-value problems for the director angle. As with the square, there are two diagonal $D$ states for which $\mathbf{n}$ in \eqref{P} is aligned along a diagonal of the rectangle, the rotated $R1$ and $R2$ states for which $\mathbf{n}$ rotates by $\pi$ radians between a pair of parallel horizontal edges and the rotated $R3$, $R4$ states for which $\mathbf{n}$ rotates by $\pi$ radians between a pair of parallel vertical edges (Fig. \ref{rectangle_D_R}). For $a >1$, the $R3$, $R4$ states have higher energies than the $R1$, $R2$ states (see \cite{tsakonas2007multistable, lewis2014colloidal} for details), breaking the energy degeneracy of the rotated solutions on a square. Interested readers are referred to \cite{fang2020surface} for bifurcation diagrams on rectangles, for different values of $a$, that capture the effects of geometrical anisotropy on NLC solution landscapes. %there are at least  the angle $\theta$ on a rectangle with various suitable Dirichlet condition. With geometrical anisotropy, there are three competing equilibria: the diagonal states for which $\mathbf{n}$ in \eqref{eq:n} is aligned along a diagonal of the rectangle, the rotated $R1$ and $R2$ states for which $\mathbf{n}$ rotates by $\pi$ radians between a pair of parallel horizontal edges and the rotated $R3$, $R4$ states for which $\mathbf{n}$ rotates by $\pi$ radians between a pair of parallel vertical edges. For $a >1$, the $R3$, $R4$ states have lower energies than the $R1$, $R2$ states (see \cite{tsakonas2007multistable, lewis2014colloidal} for details).

%The bifurcation diagrams are qualitatively similar to those reported for a square in Section \ref{sec:intro} except for the fact that the energetically expensive $R1$, $R2$ solutions are disconnected from the connected diagonal and $R3$, $R4$ solution branches. The geometrical anisotropy affects $R1$, $R2$ solution branches, in the sense that our numerical methods cannot track the pathway from the $R1$, $R2$ solutions to the unique $BD2$ solution as $\epsilon$ increases (Fig. \ref{rectangle_bifurcation}).

%\begin{figure}[b]
%    
%        \includegraphics[width=0.649\columnwidth]{./image/bifurcation_rectangle.pdf}
%        \caption{The bifurcation diagram as a function of $\epsilon$ on rectangle with $a = 1.5$. In the legend, the prefixes `s' and `u' for stable and unstable solutions respectively. Reproduced from \cite{fang2019solution} with permission from SAGE Publications}
%        \label{rectangle_bifurcation}
%\end{figure}

\section{Effects of Elastic Anisotropy}\label{sec:ani}
In this section, we study the effects of elastic anisotropy on the critical points of the reduced LdG energy (\ref{eq:reduced}) on square domains, with tangent boundary conditions. The elastic anisotropy is captured by a parameter $\hat{L}_2$ in \eqref{eq:fel}. As in previous sections, we restrict ourselves to $\Q$-tensors with three degrees of freedom $q_1$, $q_2$, and $q_3$ in \eqref{q123}. In the following paragraphs, we review the modelling details, theoretical analyses and numerical results from \cite{han+majumdar+harris+zhang+2021}.

%the equilibrium configurations on square with elastic anisotropy, including their symmetry properties in the small $\lambda$ limit. Notably, we show that the cross structure of the $WORS$ does not survive with $\hat{L}_2 \neq 0$, in the followings. 

Substituting the $\Q$-tensor ansatz \eqref{q123} into  \eqref{eq:reduced}, and writing the energy functional as a function of $(q_1,q_2,q_3)\in W^{1,2}(\Omega;\mathbb{R}^3)$ we have
\begin{align}
    J[q_1,q_2,q_3]:=&\int_{\Omega}f_{el}(q_1,q_2,q_3)+\frac{\lambda^2}{L}f_b(q_1,q_2,q_3)\,\mathrm{dA},\label{funcq123}
\end{align}
where
\begin{align}
    f_b(q_1,q_2,q_3):=&A(q_1^2+q_2^2+3q_3^2)+C(q_1^2+q_2^2+3q_3^2)^2+2Bq_3(q_1^2+q_2^2-q_3^2),
\end{align}
and
\begin{align}
    &f_{el}(q_1,q_2,q_3):=\left(1+\frac{\hat{L}_2}{2}\right)|\nabla q_1|^2+\left(1+\frac{\hat{L}_2}{2}\right)|\nabla q_2|^2+\left(3+\frac{\hat{L}_2}{2}\right)|\nabla q_3|^2 \nonumber\\
    &+\hat{L}_2(q_{1,y}q_{3,y}-q_{1,x}q_{3,x}-q_{2,y}q_{3,x}-q_{2,x}q_{3,y})+|\hat{L}_2|(q_{2,y}q_{1,x}-q_{1,y}q_{2,x}). 
\end{align}
The elastic energy density can be rewritten in the following two ways: if $\hat{L}_2\geq 0$,
\begin{gather}
    f_{el}=|\nabla q_1|^2+|\nabla q_2|^2+3|\nabla q_3|^2+\frac{\hat{L}_2}{2}((q_{1,x}+q_{2,y}-q_{3,x})^2+(q_{2,x}-q_{1,y}-q_{3,y})^2), \label{pos}
\end{gather}
    and if $\hat{L}_2<0$,
\begin{align}
    f_{el} =&(1+\hat{L}_2)(|\nabla q_1|^2+|\nabla q_2|^2+3|\nabla q_3|^2)\\
    &-\frac{\hat{L}_2}{2}((-q_{3,x}-q_{1,x}-q_{2,y})^2+(q_{2,x}-q_{1,y}+q_{3,y})^2+4|\nabla q_3|^2).\label{neg}
\end{align}
To ensure the non-negativity of the elastic energy density, we assume $\hat{L}_2\in(-1,0)$.
The corresponding EL equations are:
\begin{align}
    \left(1+\frac{\hat{L}_2}{2}\right)\Delta q_1+\frac{\hat{L}_2}{2}(q_{3,yy}-q_{3,xx})=&\frac{\lambda^2}{L}q_1(A+2Bq_3+2C(q_1^2+q_2^2+3q_3^2)), \label{q1eq} \\
    \left(1+\frac{\hat{L}_2}{2}\right)\Delta q_2-\hat{L}_2q_{3,xy}=&\frac{\lambda^2}{L}q_2(A+2Bq_3+2C(q_1^2+q_2^2+3q_3^2)), \label{q2eq}\\
    \left(1+\frac{\hat{L}_2}{6}\right)\Delta q_3+\frac{\hat{L}_2}{6}(q_{1,yy}-q_{1,xx})-\frac{\hat{L}_2}{3}q_{2,xy}=&\frac{\lambda^2}{L}q_3(A-Bq_3+2C(q_1^2+q_2^2+3q_3^2))\nonumber \\
    &+\frac{\lambda^2B}{3L}(q_1^2+q_2^2).\label{q3eq}
\end{align}
The EL equations \eqref{q1eq}-\eqref{q3eq} do not have the elegant Laplace structure and hence, are not readily amenable to analytic methods. Notably, we do not have an explicit maximum principle argument for the solutions of \eqref{q1eq}-\eqref{q3eq}, as in the Dirichlet case with $\hat{L}_2 = 0$.
\begin{figure}[b]
    
    \includegraphics[width=0.5\columnwidth]{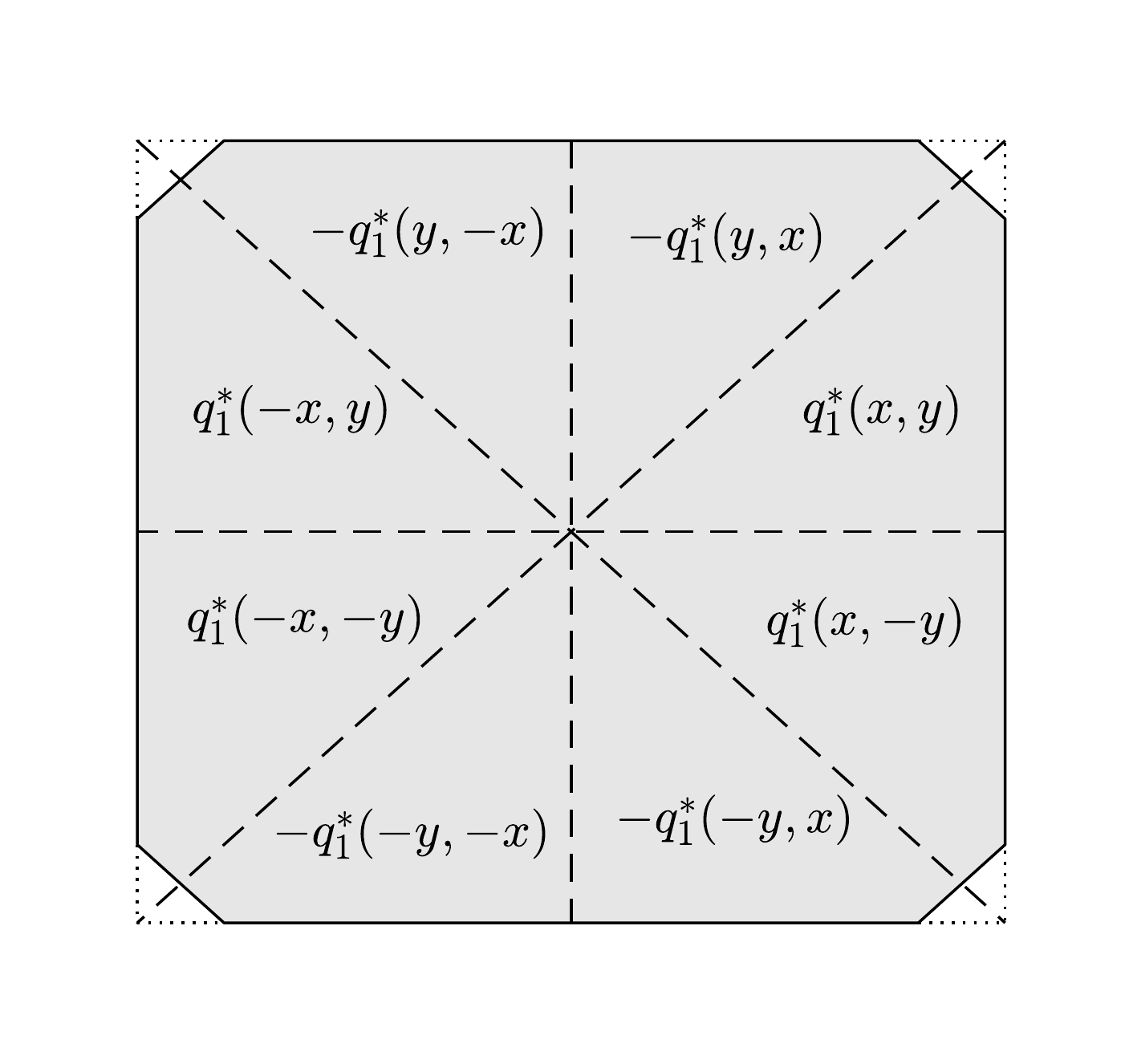}
    \caption{(Credit \cite{han+majumdar+harris+zhang+2021}) The reflected solution $q_1^s(x,y)$ in Proposition \ref{prop:forever_critical}.}
    \label{fig:reflected}
\end{figure}

 Analogous to Theorem $2.2$ in \cite{bauman2012analysis}, we can prove the existence of minimizers of $J$ in \eqref{funcq123} in the admissible class
\begin{gather} \label{Ao}
    \mathcal{A}_0:=\{(q_1,q_2,q_3)\in W^{1,2}(\Omega;\mathbb{R}^3) : q_1=q_b,\, q_2=0,\, q_3=-s_+/6\,\,\text{on}\,\,\partial\Omega\},
\end{gather}
where $q_b$ is piece-wise of class $C^1$, $q_1$ is prescribed to ensure that the tangent boundary conditions are satisfied. For $\lambda$ small enough, we can prove that the LdG energy (\ref{funcq123}) has a unique critical point but the proof is more involved than in \cite{lamy2014}, with additional embedding theorems and functional inequalities. %we use the Poincar\'{e} inequality and the relevant embedding theorem in \cite{brezis2010functional} from $W^{1,2}$ space to $L^4$ space, since $L^2$ space is not enough for the complex elastic energy density. 

\begin{figure}
    \begin{center}
    \includegraphics[width=0.9\columnwidth]{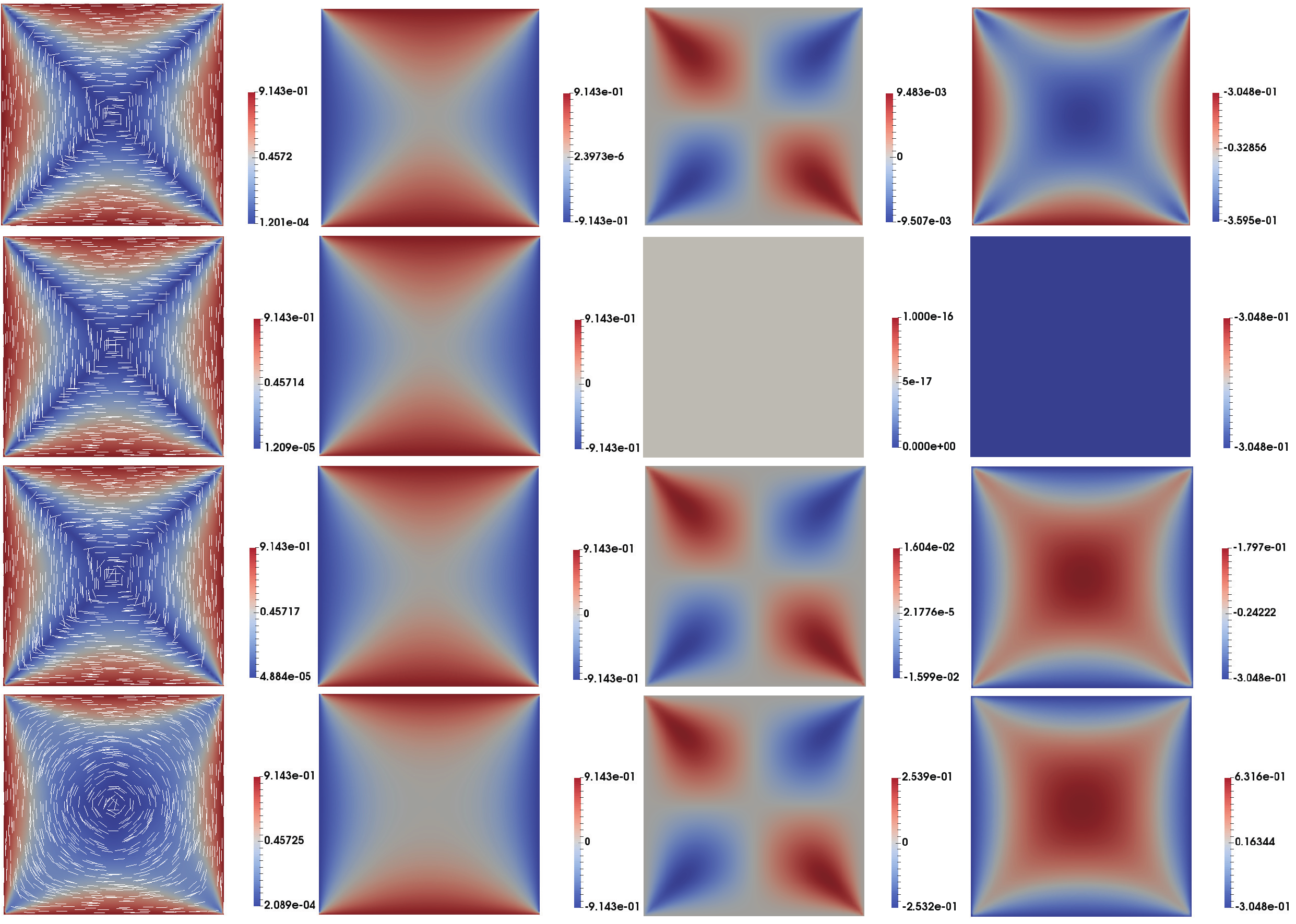}
        \caption{(Credit \cite{han+majumdar+harris+zhang+2021}) The unique stable solution of the Euler-Lagrange equations \eqref{q1eq}--\eqref{q3eq}, with $\bar{\lambda}^2 = 5$, and (from the first to fourth row) $\hat{L}_2 = -0.5$, $0$, $1$ and $10$, respectively. In the first column, we plot the $(q_1,q_2)$ profile. We plot the corresponding $q_1,q_2$ and $q_3$ profiles, in the second to fourth columns, respectively.}
        \label{fig:RING_lambda_5}
    \end{center}
\end{figure}

%Using the symmetry property of boundary condition, geometry and the structure of EL equations in \eqref{q1eq}-\eqref{q3eq}. 
We can analytically construct a symmetric critical point for all admissible values of $\hat{L}_2$ and edge lengths $\lambda$, and we quote the relevant proposition from \cite{han+majumdar+harris+zhang+2021} below.
\begin{proposition} \label{prop:forever_critical}
    There exists a critical point $(q_1^s, q_2^s, q_3^s)$ of the energy functional (\ref{funcq123}) in the admissible space $\mathcal{A}_0$, for all $\lambda>0$, %with the symmetry property 
    such that $q_1^s$ is odd about the square diagonals and $x$- and $y$-axis (see Fig. \ref{fig:reflected}), $q_2^s$ has even reflections about the square diagonals and odd reflection about $x$- and $y$-axis, $q_3^s$ has even reflections about the square diagonals and $x$- and $y$-axis. Subsequently, $q_1:\Omega\to\mathbb{R}$ vanishes along the square diagonals $y=x$ and $y=-x$, and the function $q_2:\Omega\to\mathbb{R}$ vanishes along $y=0$ and $x=0$.
\end{proposition}
%the has odd reflections about the square diagonals and odd reflection about $x$- and $y$-axis (see Fig. \ref{fig:reflected}), $q_2^s$ has even reflections about the square diagonals and odd reflection about $x$- and $y$-axis, $q_3^s$ has even reflections about the square diagonals and $x$- and $y$-axis. Subsequently, $q_1:\Omega\to\mathbb{R}$ vanishes along the square diagonals $y=x$ and $y=-x$, and the function $q_2:\Omega\to\mathbb{R}$ vanishes along $y=0$ and $x=0$.
%\end{proposition}
Subsequently, we can exploit the structure of the equations \eqref{q1eq}-\eqref{q3eq} and the boundary conditions to prove that for $A<0$ and $\hat{L}_2\neq0$, the critical point constructed in Proposition \ref{prop:forever_critical}, %denoted by $(q_1^s, q_2^s, q_3^s)$, 
has non-constant $q_2^s$ on $\Omega$, for all $\lambda>0$.
This symmetric critical point is globally stable for small domains size, i.e., the edge length $\lambda$ is small enough (see Fig. \ref{fig:RING_lambda_5}).
 %All of them satisfy the symmetry property demonstrated in Proposition \ref{prop:forever_critical} and one can check that $q_1:\Omega\to\mathbb{R}$ vanishes along the square diagonals $y=x$ and $y=-x$, and the function $q_2:\Omega\to\mathbb{R}$ vanishes along $y=0$ and $x=0$. 
 When $\hat{L}_2 = 0$, this symmetric critical point is the $WORS$ defined by (\ref{WQ}). Notably, $q_2=0$ everywhere for the $WORS$ (refer to (\ref{q123})), which is equivalent to having a set of constant eigenvectors in the plane of $\Omega$. When $|\hat{L}_2|>0$, $q_2$ and $q_3$ are non-constant, which means we lose the constant eigenvectors and subsequently the cross structure in $WORS$. When $\hat{L}_2=-0.5,1,$ and $10$, we have a central $+1$-point defect in the profile of $(q_1,q_2)$, and we label this as the $Ring^+$ solution (Fig. \ref{fig:RING_lambda_5}). %Hence, we lose the $WORS$ solution on square domains with $\hat{L}_2\neq 0$.% This proof indicates that for $\hat{L}_2\neq 0$ and small enough $\lambda$ we lose the fixed eigen-frame in $WORS$ state.

We investigate the effect of $\hat{L}_2$ on the $WORS$-profile, using asymptotic methods. % unique states when $\lambda\to 0$ by an asymptotic study based on the $WORS$ solution. 
The $WORS$ in the form of \eqref{q123}, given by the triplet $(q,0,-B/6C)$ at the fixed temperature $A = -B^2/3C$, where $q$ is a solution of the Allen-Cahn equation, as in \cite{canevari2017order}. 
%:
%\begin{gather}
%\Delta q=\frac{2C\lambda^2}{L}q\left(q^2-\frac{B^2}{4C^2}\right), \qquad q=q_{b}\quad\text{on}\quad\partial\Omega.  %\label{WORS}
%\end{gather}
With the leading order approximation given by $(q,0,-B/6C)$, we expand $q_1, q_2, q_3$ in powers of $\hat{L}_2$ as follows:
\begin{equation}
\begin{aligned}\label{L2_0_asymptotics}
q_1(x,y)&=q(x,y)+\hat{L}_2f(x,y)+\dots \\
q_2(x,y)&=\hat{L}_2g(x,y)+\dots  \\
q_3(x,y)&=-\frac{B}{6C}+\hat{L}_2h(x,y)+\dots     
\end{aligned}
\end{equation}
for some functions $f, g, h$ which vanish on the boundary. 

For $\lambda$ small enough, one can show that the corrections $(f,g,h)$ are unique, $g\equiv0$ on $\Omega$ and $f(x,y) = 0$ on diagonals. %\textcolor{red}{(Fig. \ref{fig:L2_0_asymptotic_compare})}. 
Hence, for $\lambda$ small enough, the cross structure of the $WORS$ is lost mainly because of effects of $\hat{L}_2$ on the component $q_3$. %by following the approach in \cite{lamy2014} to show that $F$ is strictly convex in $W^{1,2}_0(\Omega;\mathbb{R}^3)$ for sufficiently small $\lambda$. Hence, for $\lambda$ small enough, we have $g\equiv0$ on $\Omega$.
%Using the symmetry property of equations in \eqref{feq}-\eqref{heq},
% we can check that if $(q(x,y),f(x,y),g(x,y),h(x,y))$ is a solution of \eqref{WORS},\eqref{feq}--\eqref{heq}, then the quadruplets, $(-q(y,x),-f(y,x),g(y,x),h(y,x))$ and $(q(-x,y),f(-x,y),g(-x,y),h(-x,y))$, are also solutions of \eqref{WORS},\eqref{feq}--\eqref{heq}. Thus we have $f(x,y) = 0$ on diagonals for $\lambda$ small enough. Hence, for $\lambda$ small enough, the cross structure of the $WORS$ is lost mainly because of effects of $\hat{L}_2$ on the component $q_3$.

%At $\hat{L}_2 = 1000$, there are one -1 central point defect and four +1/2 point defects. The reason for this complicated profile is unknown.
\begin{figure}
    \begin{center}
    \includegraphics[width=0.9\columnwidth]{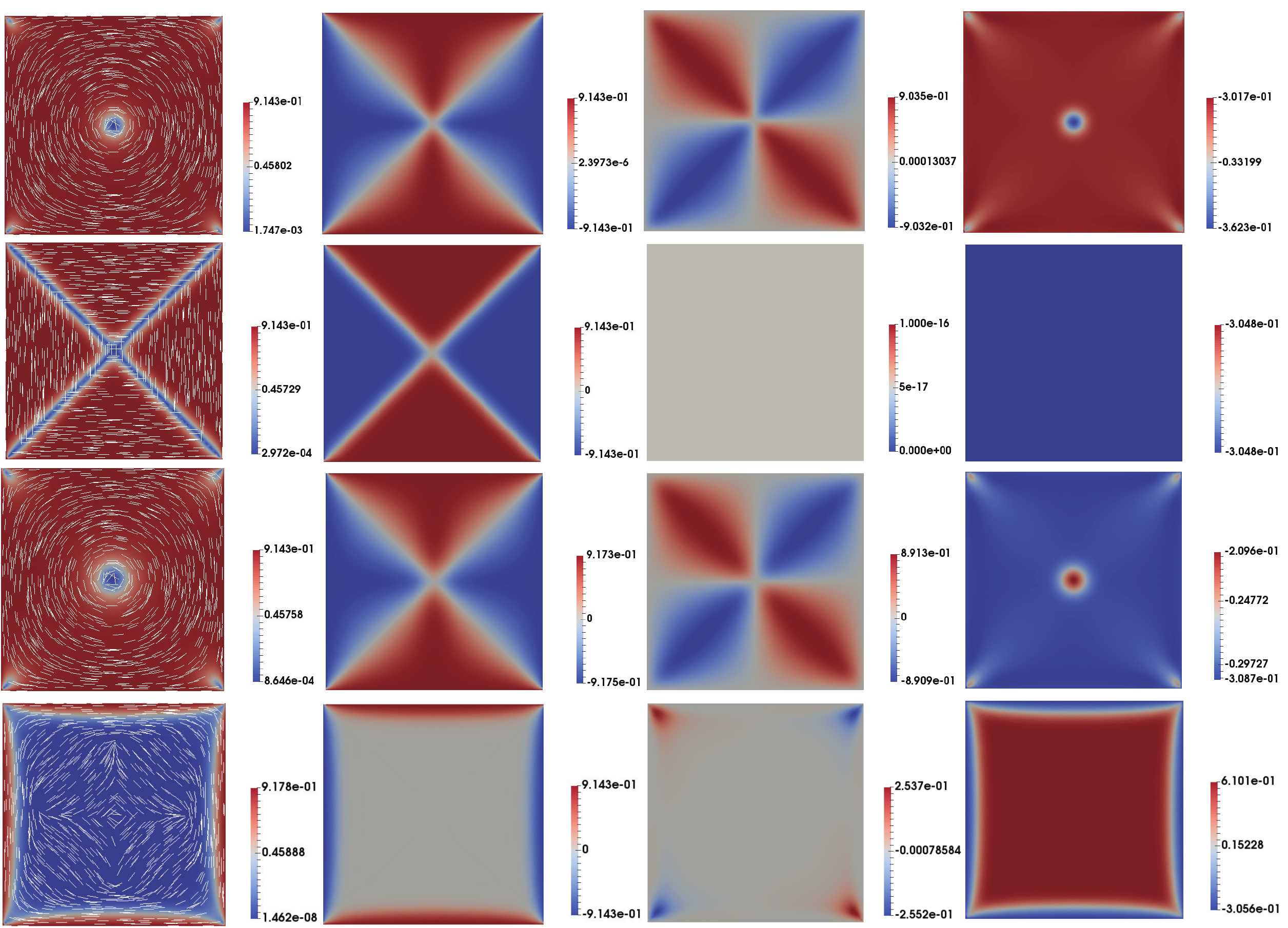}
        \caption{(Credit \cite{han+majumdar+harris+zhang+2021}) A solution branch for the Euler-Lagrange system \eqref{q1eq}--\eqref{q3eq} with $\bar{\lambda}^2 = 500$, and $\hat{L}_2 =-0.5$, $0$, $1$ and $10$ respectively, plotted in the first to fourth row respectively. This solution branch is a symmetric solution branch, as described in Proposition~\ref{prop:forever_critical}. When $\hat{L}_2 = -0.5$, $0$ and $1$, the plotted solution is unstable. When $\hat{L}_2 = 10$, the plotted solution is stable. The first column contains plots of $(q_1,q_2)$. In the second to fourth column, we plot the corresponding, $q_1,q_2$ and $q_3$ profiles.}
        \label{fig:RING_lambda_500}
    \end{center}
\end{figure}
 We work at the fixed temperature $A = -B^2/3C$ for all the following numerical results. We perform a parameter sweep of $\bar{\lambda}^2$, from $5$ to $500$, and find one of the symmetric solution branches constructed in Proposition \ref{prop:forever_critical}, with various fixed $\hat{L}_2$. The solutions with $\bar{\lambda}^2=500$ are plotted in Fig. \ref{fig:RING_lambda_500}. When $\hat{L}_2=0$, we recover the familiar $WORS$ for all $\lambda>0$. %Notably, when $\hat{L}_2\neq 0$, we lose the defective cross structure on the diagonals of square which means we the $WORS$ state does not survive. 
When $-1<\hat{L}_2<0$, the solution exhibits a $+1$-defect at the square center, %continued from the $Ring^+$ branch and hence, 
 and we refer to it as the $Ring^+$ solution. When $\hat{L}_2$ is positive and moderate in value, we recover the $Ring^+$ solution branch %and the corresponding $q_3<-s_+/6$ at the square center for negative $\hat{L}_2$, but  
and $q_3>-s_+/6$ at the square centre. When $\hat{L}_2$ is large enough, we discover a new symmetric solution which is approximately constant,  $(q_1, q_2, q_3) = (0,0,s_+/3)$, away from the square edges, as shown in the fourth row of Fig. \ref{fig:RING_lambda_500} for $\hat{L}_2=10$. We refer to this novel solution as the \emph{Constant} solution.

As stated in Section \ref{sec:ben}, for large $\lambda$ and with $\hat{L}_2 = 0$, the $D$ and $R$ states are the competing energy minimizers in this reduced framework. %stable states .
For large $\lambda$, with small or moderate $\hat{L}_2$, the $D$ and $R$ states still survive. %The effect of the elastic anisotropy concentrates near the square vertices (Fig. \ref{fig:D}). 
When $\hat{L}_2 = 0$ and for fixed $A<0$,  $s^2 = q_1^2+q_2^2\approx s_+^2/4$, $q_3= -s_+/6$ almost everywhere on $\Omega$. As $|\hat{L}_2|$ increases, $q_3$ deviates significantly from the limiting value $q_3^\infty= -s_+/6$, near the square vertices; the deviation being more significant near the bend vertices compared to the splay vertices. %Notably, the value of $q_3$ near the vertices increases as $\hat{L}_2$ increases and, 
From an optical perspective, we expect to observe larger defects near the square vertices for anisotropic materials with $\hat{L}_2 \gg 1$. 
%\begin{figure}
    %\begin{center}
   % \includegraphics[width=0.85\columnwidth]{./image/D_R.pdf}
   % \caption{(Credit \cite{han+majumdar+harris+zhang+2021}) The $D$ and $R$ solutions of the Euler-Lagrange equation \eqref{q1eq}--\eqref{q3eq} with $\hat{L}_2 = 3.5$ and $\bar{\lambda}^2= 1000$. In the first and third columns, we plot the $(q_1,q_2)$ profile. The second and fourth columns are the corresponding plots of $q_3$. We see that $q_3$ is approximately constant on $\Omega$, except for near the vertices.}
   % \label{fig:D_R}
    %\end{center}
%\end{figure}
%\begin{figure}
 %   \begin{center}
  %  \includegraphics[width=\columnwidth]{./image/D_L2_10_30_45%%.pdf}
    %\caption{The $D$ solution of the equations, \eqref{q1eq}--\eqref{q3eq}, with $\bar{\lambda}^2= 1000$, with $\hat{L}_2 = -0.5, 0, 10,30$ and $45$, respectively. The effects of increasing $\hat{L}_2$ are most pronounced near the bend vertices.}
    %\label{fig:D}
   % \end{center}   
%\end{figure}
%We take the limiting profile for $D$ and $R$ as initial condition to obtain the corresponding solution of \eqref{q1eq}--\eqref{q3eq} with $\hat{L}_2 = 3.5$ and $\frac{\lambda^2}{2C} = 1000$ by using Newton's method for the Euler-Lagrange equations \eqref{q1eq}--\eqref{q3eq}. The $D$ and $R$ stable states are shown in Fig. \ref{fig:D_R}. 
For large $\lambda$ and large $\hat{L}_2$, the $D$, $R$ and $Constant$ states are three energetically competing states.
In \cite{lewis2014colloidal}, as $\lambda\to\infty$, the authors compute the limiting energies of $D$ and $R$ solutions, and the energy estimates are linear in $\hat{L}_2$.
The $Constant$ solution has transition layers near the square edges %on the boundary from $(0,0,s_+/3)$ to $(s_+/2,0,-s_+/6)$ or $(- s_+/2,0,-s_+/6)$.
and as in Section 4 of \cite{wang2019order}, by using the geodesic distance theory, we %calculate the transition costs of $Constant$ which is independent of $\hat{L}_2$. Hence, there is 
can show that there is a critical value $\hat{L}_2^*$, such that for $\hat{L}_2>\hat{L}_2^*$, the limiting $Constant$ solution has lower energy than the competing $D$ and $R$ solutions.

\begin{figure}
\centering
    \begin{subfigure}{0.49\textwidth}
        \centering
        \includegraphics[width=\columnwidth]{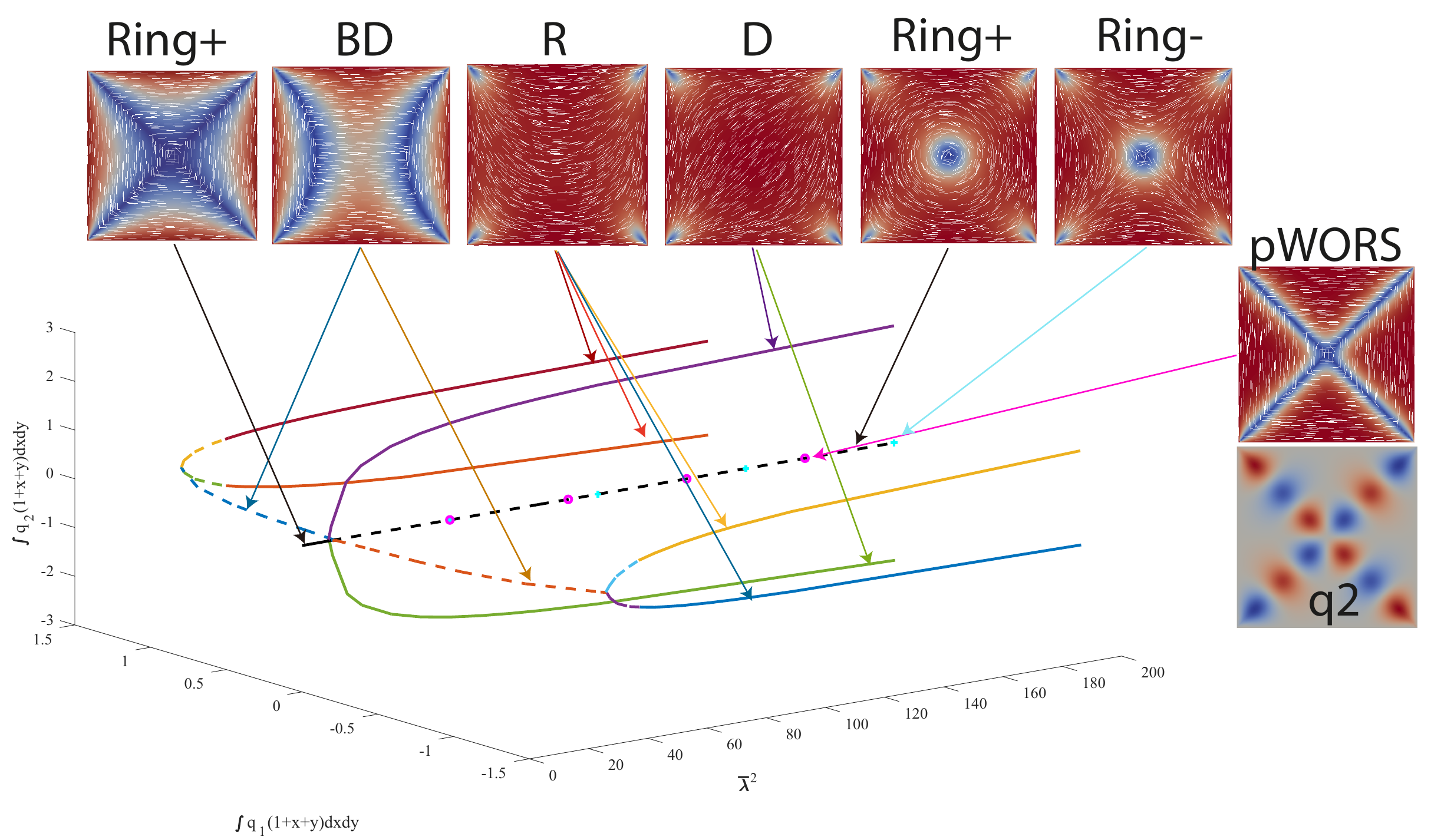}
    \end{subfigure}
    \begin{subfigure}{0.49\textwidth}
        \centering
        \includegraphics[width=\columnwidth]{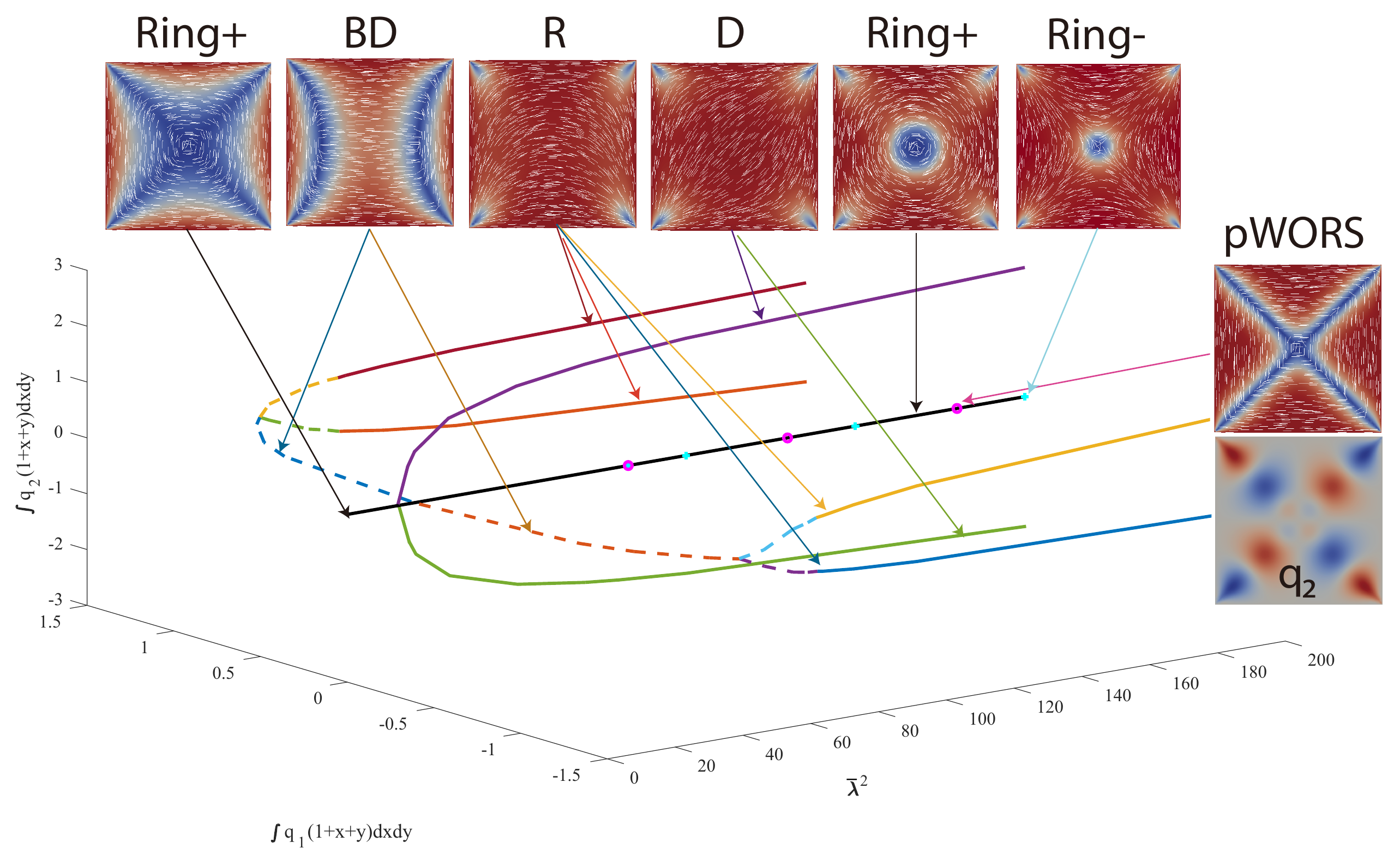}
    \end{subfigure}
    \begin{subfigure}{0.49\textwidth}
        \centering
        \includegraphics[width=\columnwidth]{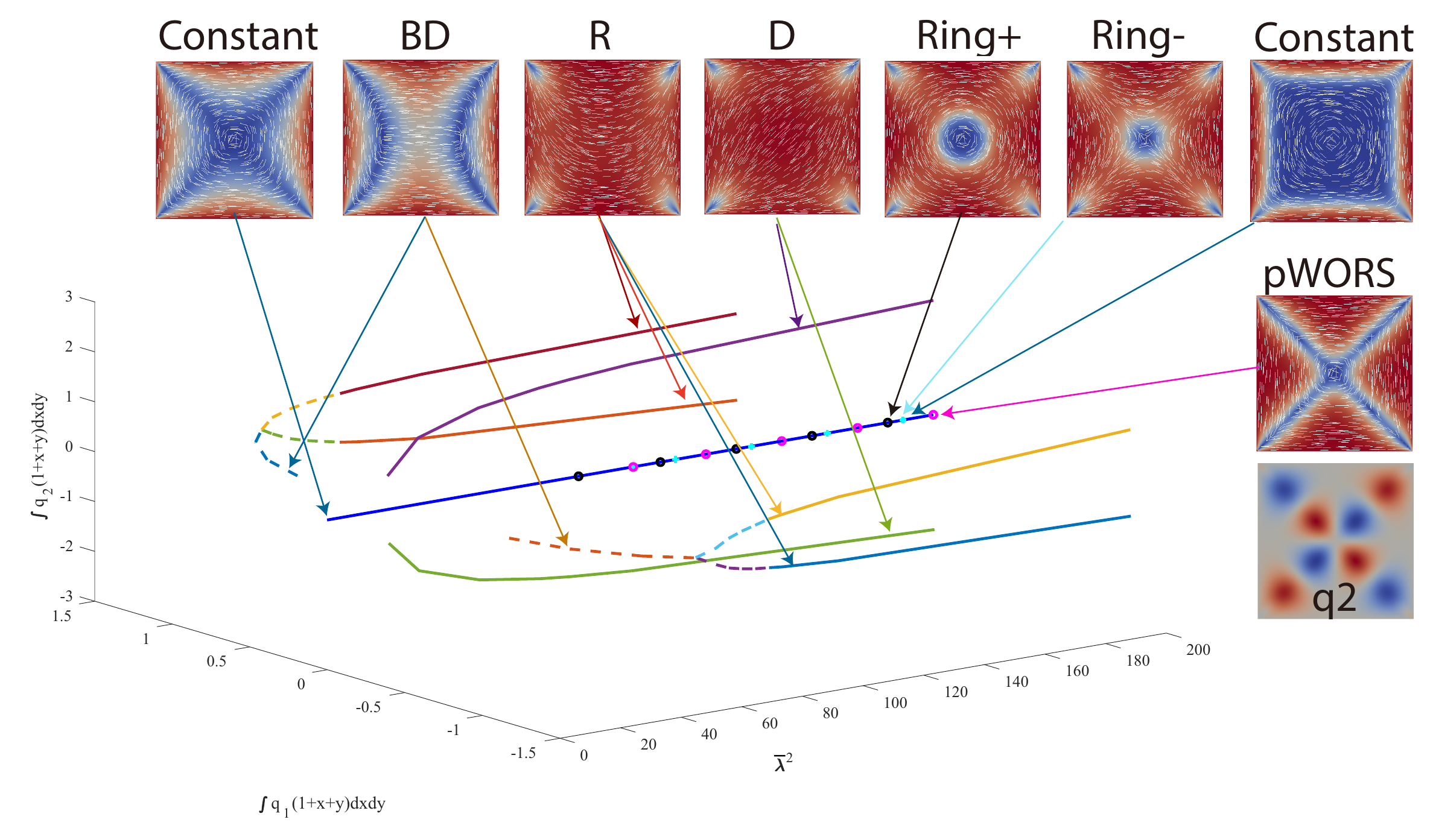}
    \end{subfigure}
    \begin{subfigure}{0.49\textwidth}
        \centering
        \includegraphics[width=0.9\columnwidth]{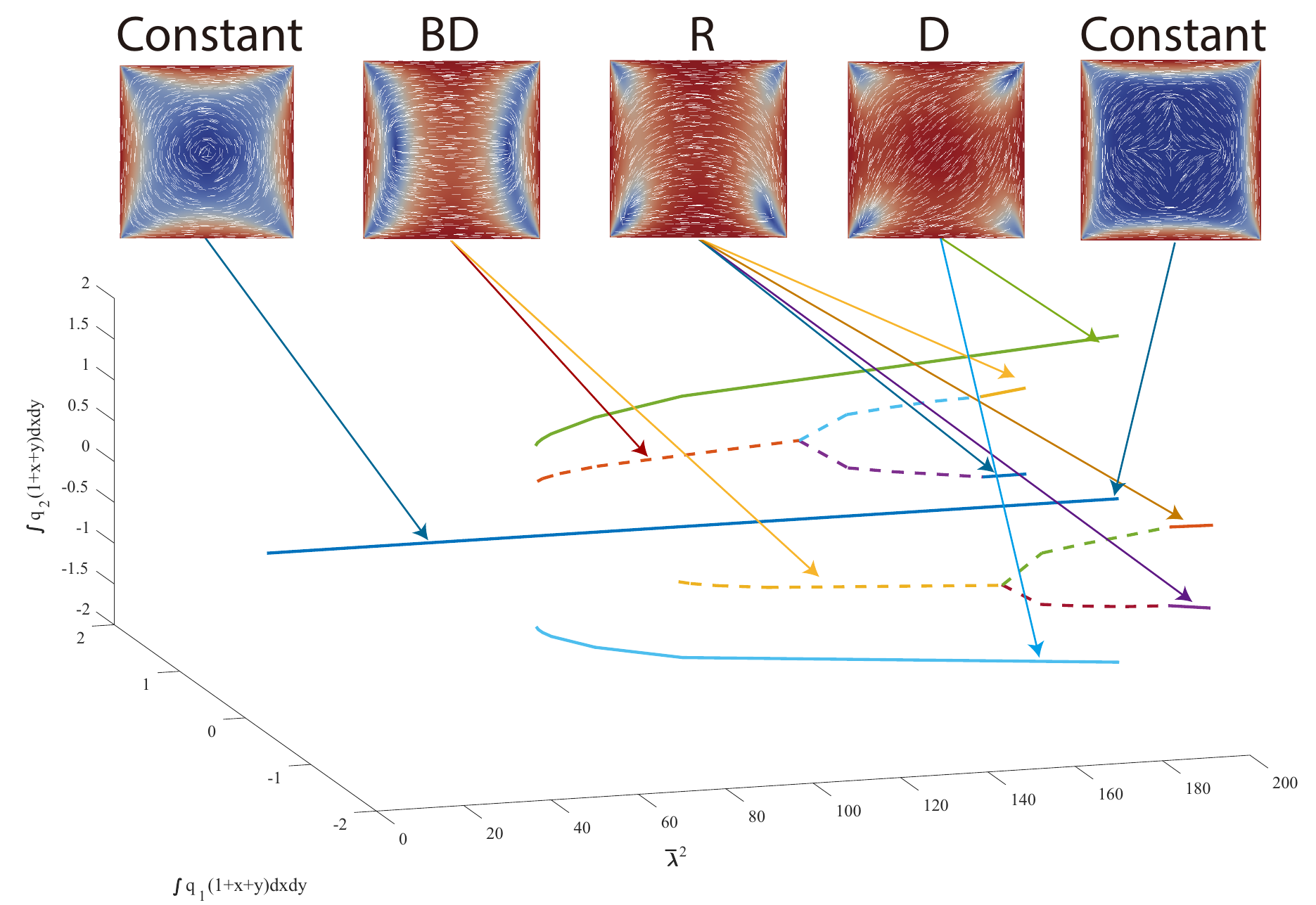}
    \end{subfigure}
        \caption{(Credit \cite{han+majumdar+harris+zhang+2021}) Bifurcation diagrams for the LdG model in square domain with $\hat{L}_2 = 1$, $2.6$, $3$, and $10$ in top left, top right, bottom left and bottom right.}
        \label{fig:bifurcation_diagram_ani}
\end{figure}

In what follows, we compute bifurcation diagrams as a function of $\lambda$, with fixed temperature $A = -B^2/3C$, for five different values of $\hat{L}_2 = 1, 2.6, 3, 10$ in Fig. \ref{fig:bifurcation_diagram_ani}. %We will use these diagrams to infer qualitative solution trends in terms of the edge length, $\lambda$. 
We numerically discover at least 5 classes of symmetric critical points constructed in Proposition~\ref{symmetry_proposition} --- the $WORS$, $Ring^\pm$, $Constant$ and the $pWORS$ solutions, of which the $WORS$, $Ring+$ and the $Constant$ solutions can be stable.
The bifurcation diagram for $\hat{L}_2 =0$, the elastically isotropic case, is discussed in Section \ref{sec:ben}. %, where the unique solution $WORS$ for small $\lambda$ bifurcates into stable $D$ and into stable $R$ through unstable $BD$ state.
For $\hat{L}_2 = 1$,  the $WORS$ ceases to exist and the unique solution in the $\lambda \to 0$ limit, is the stable $Ring^+$ solution. At the first bifurcation point $\lambda = \lambda^*$, the $Ring^+$ solution bifurcates into an unstable $Ring^+$ and two stable $D$ solutions. At the second bifurcation point, $\lambda = \lambda^{**}>\lambda^*$, the unstable $Ring^+$ bifurcates into two unstable $BD$ solutions and for $\lambda = \lambda^{***}>\lambda^{**}$, the unstable $Ring^-$ and unstable $pWORS$ solution branches appear. In the $(q_1,q_2)$ plane, the $pWORS$ has a constant set of eigenvectors away from the diagonals, and has multiple $\pm 1/2$-point defects on the two diagonals, so that the $pWORS$ is similar to the $WORS$ away from the square diagonals. The $Ring^-$ and $pWORS$ are always unstable and the $Ring^+$ solution has slightly lower energy than the $Ring^-$. The unstable $pWORS$ has higher energy than the unstable $Ring^\pm$ solutions when $\lambda$ is large. 
%As $\hat{L}_2\geq2.6$, the $Ring^+$ solution is always stable, at least when $\lambda\leq 200$.
The solution landscape for $\hat{L}_2=1$ and $\hat{L}_2=2.6$ are qualitatively similar.
% but for $\hat{L}_2=2.6$, the $Ring^+$ solution is stable for $\bar{\lambda}^2\leq 200$ and the unstable $pWORS$ and $Ring^-$ appear for large $\lambda$.  
For $\hat{L}_2 = 3$, the unique stable solution, for small $\lambda$, is the $Constant$ solution, which remains stable for $\bar{\lambda}^2\leq 200$. The $Constant$ solution approaches $(q_1,q_2, q_3) \to (0,0, s_+/3)$ as $\lambda$ or $\hat{L}_2$ gets large. The $BD$ and $D$ solution branches
are disconnected from the stable $Constant$ solution branch. For $\lambda = \lambda^*$, the stable $Ring^+$ appears and for $\lambda = \lambda^{**}>\lambda^*$, the unstable $Ring^-$ and $pWORS$ appear. 
%The $pWORS$ and $Ring^\pm$ exist for moderately small values of $\hat{L}_2$ is small, and $\lambda$ is large.
For $\hat{L}_2 = 10$, i.e., for very anisotropic materials, the $pWORS$ and $Ring^\pm$ states disappear, and the $Constant$ solution does not bifurcate to any known states. 
The $Constant$ solution has lower energy than the $R$ and $D$ solutions for large $\lambda$. 
For much larger values of $\hat{L}_2$, we only numerically observe the $Constant$ solution branch, for the numerically accessible values of $\lambda$.

To summarise, the primary effect of the anisotropy parameter, $\hat{L}_2$, is on the unique stable solution for small $\lambda$. The elastic anisotropy destroys the cross structure of the $WORS$, and also enhances the stability of the $Ring^+$ and $Constant$ solutions. In fact, the $Constant$-solution is only observed for large $\hat{L}_2$. A further interesting feature for large $\hat{L}_2$, is the disconnectedness of the $D$ and $R$ solution branches from the parent $Constant$ solution branch. This indicates novel hidden solutions for large $\hat{L}_2$, which may have different structural profiles to the discussed solution branches. 

\section{NLC Solution Landscapes on a Hexagon}\label{sec:SL}
We have studied NLC equilibria on regular polygons, with or without elastic anisotropy. In this section, we investigate the solution landscape of a thin layer of NLC on a 2D hexagon, including stable and unstable critical points of the reduced LdG energy \eqref{eq:reduced}. % which includes unstable states such as $BD$ and transition pathways between stable states.
The hexagon is a generic example of a 2D polygon with an even number of sides: the hexagon supports the generic $Ring$  solution for small domains, does not support the special symmetric solutions exclusive to a square (constructed in Proposition~\ref{symmetry_proposition}) and is better suited to capture generic trends with respect to geometrical parameters, as illustrated in Section \ref{sec:pol}.

Firstly, we recap the essential concepts of a solution landscape. A \emph{Solution Landscape} is a pathway map of connected solutions of a system of partial differential equations, in this case the Euler-Lagrange equations of the reduced LdG energy in (\ref{eq:reduced}). The solution landscape starts at a parent state (typically an unstable critical point of the LdG energy), and connects to stable energy minimizers via intermediate unstable critical points. More precisely, we can measure the degree of instability of an unstable critical point by means of its Morse index \cite{milnor1969morse}. % and connecting to admissible stable energy minima with zero Morse index, via intermediate saddle points and transition states.
The Morse index of a critical/stationary point of the free energy, is the number of negative eigenvalues of its Hessian matrix \cite{milnor1969morse}. 
Energy minima or experimentally observable stable states are index-$0$ stationary points of the free energy with no unstable directions. 
A confined NLC system can switch between different energy minima or stable states, by means of an external field, thermal fluctuations, and mechanical perturbations.
The switching requires the system to cross an energy barrier separating the two stable states, with an intermediate transition state.
The transition state is an index-$1$ saddle point, the highest energy state along the transition pathway connecting the two stable states \cite{zhang2016recent}. There are typically multiple transition pathways, with distinct transition states, and the optimal transition pathway has the lowest energy barrier.
The reader is referred to \cite{kusumaatmaja2015free} for transition pathways on a square domain with tangent boundary conditions and to \cite{han2019transition} for transition pathways on a cylindrical domain with homeotropic/normal boundary conditions. Transition states are the simplest kind of saddle points of the free energy. Besides stable states and transition states, there are high-index saddle points with highly symmetric profiles and multiple interior defects, all of which offer fundamentally new scientific prospects.
 
\begin{figure}
\centering
\includegraphics[width=0.8\columnwidth]{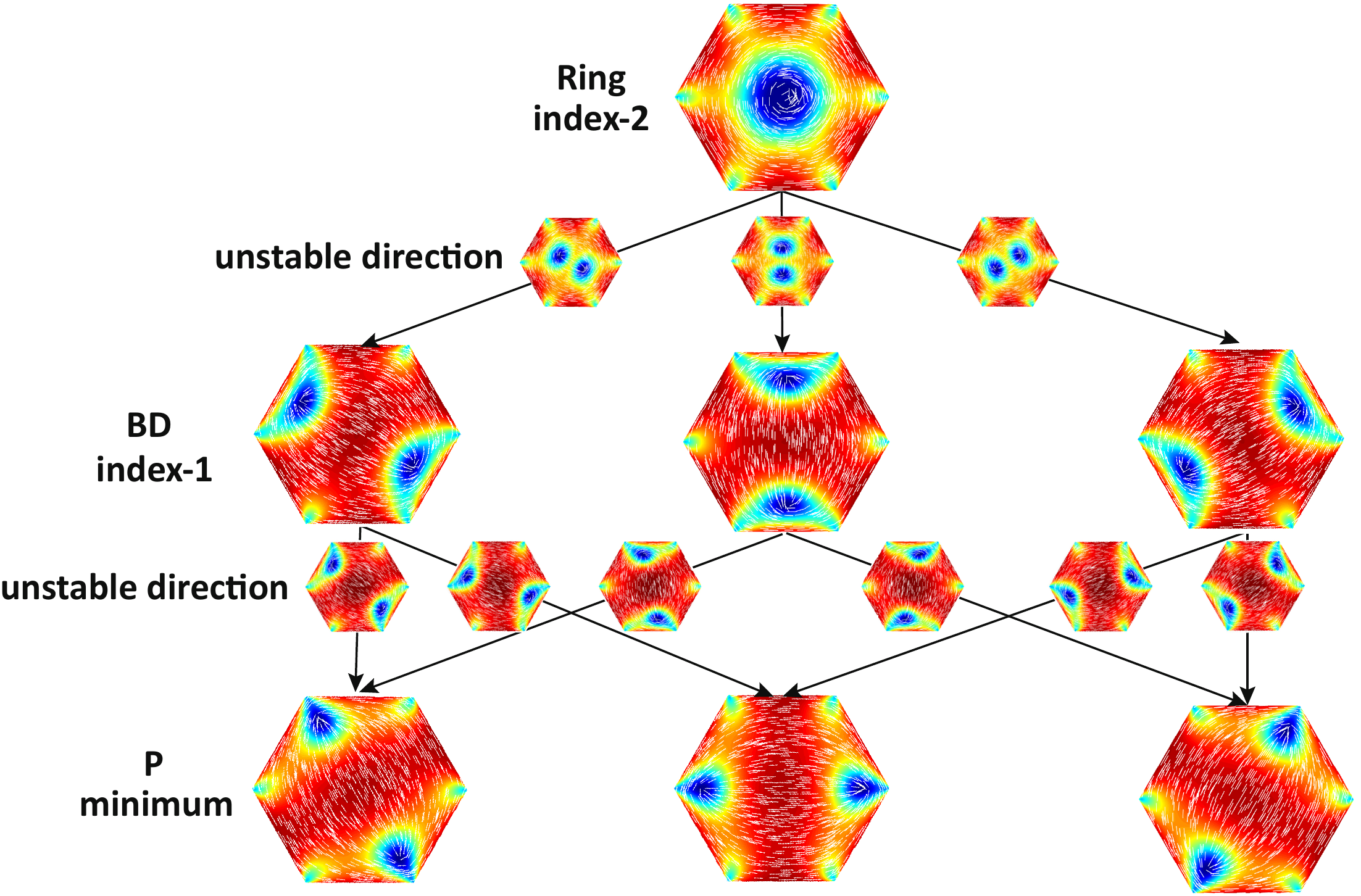}
\caption{Solution landscape at $\bar{\lambda}^2=70$. 
The index-$2$ $Ring$  is the parent state and connects to three index-$1$ $BD$ solutions along its unstable directions. 
Each $BD$ solution connects to two $P$ minima along $BD$'s single unstable direction.
%The color code is related to $|\P|^2$ and the blue regions are defects. 
%The white lines correspond to the planar nematic director without an arrow, since $\n \equiv -\n$ for nematics.
Reproduced from \cite{han2020SL} with permission from IOP Publishing and the London Mathematical Society}
\label{figure:70}
\end{figure}

In Section \ref{sec:pol}, we have reviewed the typical solutions, including $Ring$, $BD$, $Para$($P$), and $Meta$($M$), and the bifurcation diagram (Fig. \ref{hexagon_pentagon_bifurcation_diagram}(a)) of the critical points of (\ref{eq:reduced}), on a 2D hexagon. 
In what follows, we review results from \cite{han2020SL} for NLC solution landscapes on regular 2D hexagons, as a function of the hexagon edge length, $\lambda$, at the fixed temperature, $A = -B^2/3C$. % the solution landscape on hexagon with small and large domain size. 
When $\lambda^2$ is sufficiently small, the $Ring$ solution is the unique stable solution as stated in Section \ref{sec:pol}.
For $\bar{\lambda}^2 \approx 10$, the $Ring$  solution transitions from being a zero-index solution to an index-$2$ saddle point solution (with two equal negative eigenvalues), and we additionally have index-$1$ $BD$ solutions and the index-$0$ $P$ solutions.
The solution landscape for $\bar{\lambda}^2 =70$ is illustrated in Fig. \ref{figure:70}, showing the relationships between $Ring$, $BD$, and $P$ solutions.
The $Ring$ solution is the parent state, i.e., the highest-index saddle point solution.
Following each unstable eigen-direction of the $Ring$ solution shown in Fig. \ref{figure:70}, the central $+1$ point defect splits into two defects that relax around a pair of opposite edges, i.e., the $BD$ solutions.
The two $BD$ defects move from opposite edges to opposite vertices, following the single unstable eigenvector of the $BD$ solution and converging to the corresponding $P$ solution.

\begin{figure}
\begin{center}
\includegraphics[width=0.75\columnwidth]{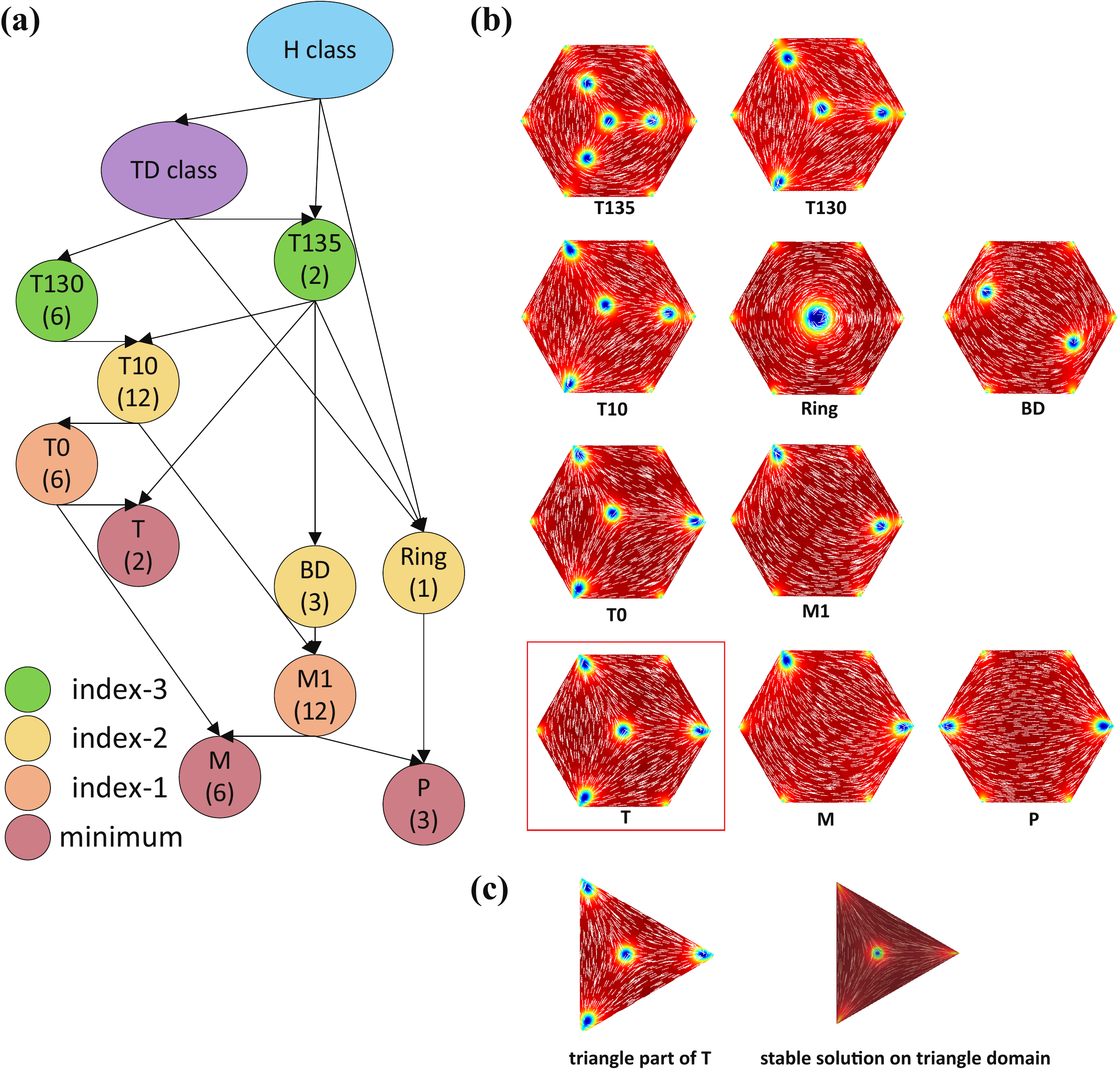}
\caption{(a) Solution landscape at $\bar{\lambda}^2 = 600$. 
(b) The configurations corresponding to (a). 
(c) The triangle part of $T$ solution on a hexagonal domain $\Omega$ and stable $Ring$  solution on a triangle domain with $\bar{\lambda}^2 = 450$. Reproduced from \cite{han2020SL} with permission from IOP Publishing and the London Mathematical Society}
\label{figure:600}
\end{center}
\end{figure}

The solution landscape is quite complicated for $\bar{\lambda}^2=600$, as shown in Fig. \ref{figure:600}(a). 
There are three notable numerical findings in this regime: 
a new stable $T$ solution with an interior $-1/2$ defect; 
new classes of saddle point solutions, $H$ and $TD$, with high symmetry and high indices; new saddle points with asymmetric defect locations. 

The stable index-$0$ $T$ solution is our first stable solution with an interior $-1/2$ defect at the centre of the hexagon, for $\bar{\lambda}^2> 250$. 
The competing stable states, $P$ and $M$, have defects pinned to the vertices, and these vertex defects are a natural consequence of the tangent boundary conditions and topological considerations (the total topological degree of the boundary condition is zero). 
%We speculate that there may be other stable solutions with interior point defects, particularly on polygons with a greater number of sides, since the disc has stable planar polar solutions with two interior $+1/2$ defects. 
%We also remark that 
The $T$ solution on a hexagon (for large $\lambda$) is strongly reminiscent of the $Ring$  solution on a regular triangle (Fig. \ref{figure:600}(c)), as reported in Section \ref{sec:pol}, suggesting that we can build new solutions by tessellating solutions on simpler building block-type polygons, such as the triangle and the square.

\begin{figure}
\begin{center}
\includegraphics[width=\columnwidth]{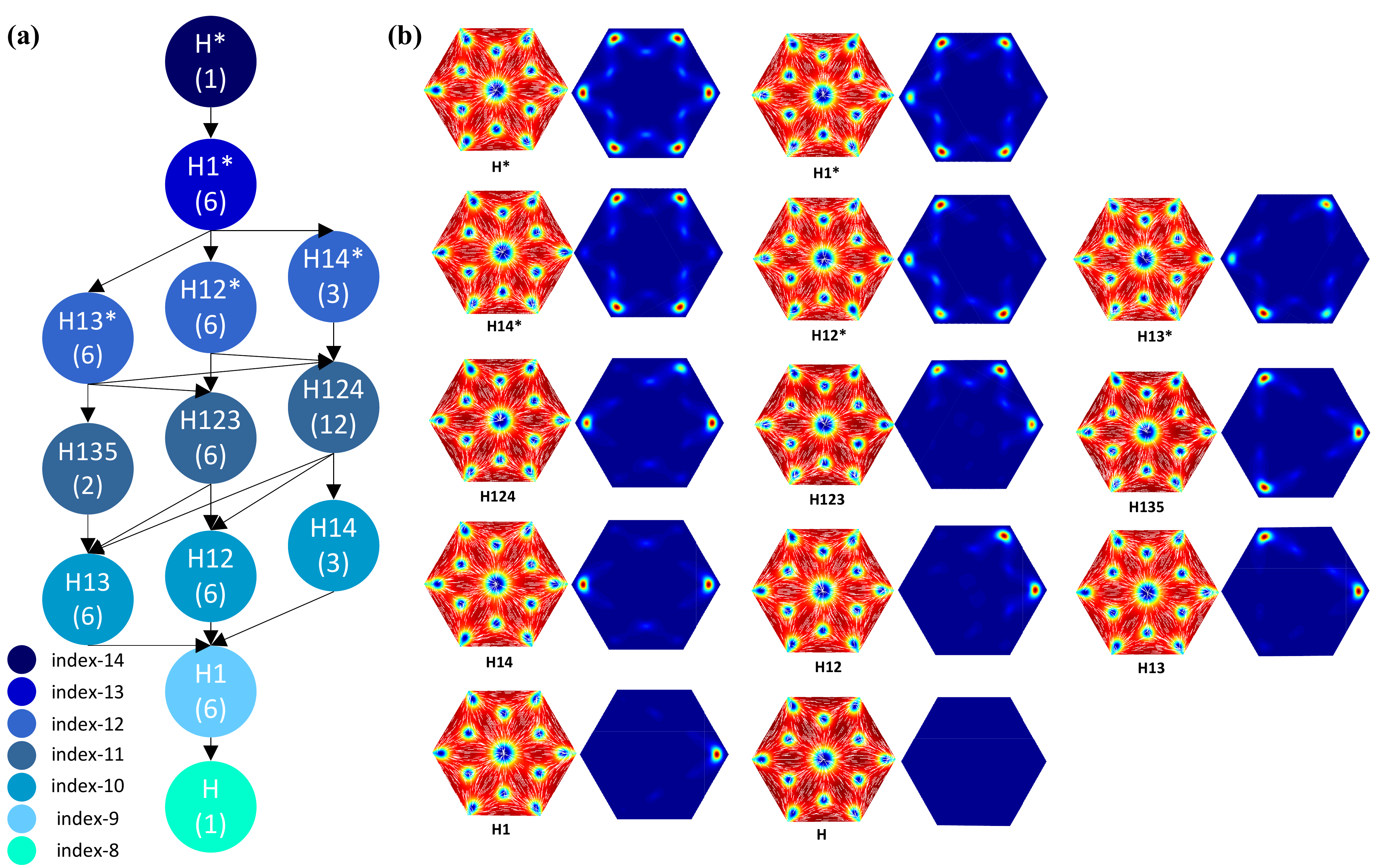}
\caption{(a) Solution landscape of the $H$ class. 
(b) The corresponding configurations and plots of $|\P-\P^H|$, where $\P$ is any solution in the $H$ class, and $\P^H$ is the index-$8$ $H$ solution. Reproduced from \cite{han2020SL} with permission from IOP Publishing and the London Mathematical Society}
\label{figure:H}
\end{center}
\end{figure}

We numerically find a new class of saddle point solutions with high Morse indices and multiple interior defects, labelled as $H$-class solutions, which have Morse indices ranging from $8$ to $14$ (Fig. \ref{figure:H}). Notably, the parent state is the index-$14$ $H*$ saddle point solution connecting to the lowest index-$8$ saddle point solution, labelled as $H$. 
%Both of these states belong to the symmetry group $G_6:=\{S\in O(2):S\Omega \in \Omega\}$ (same as the $Ring$  solution). The difference between states in $H$-class is the location and number of splay-like (bend-like) vertices.  
%We illustrate the subtle differences by plotting $|\P-\P^H|$, where $\P$ is a solution of Eq. \ref{Euler_Lagrange} in $H$-class and $\P^H$ is the index-$8$ $H$ solution. 
%The differences concentrate on the vertices with conspicuous red or white points in the dark blue background (Fig. \ref{figure:H}(b)). 
%These conspicuous points are localised near or at the bend-like vertices.
%The subscript $*$ labels the splay-like vertices (complement of bend-like vertices)
The saddle point $H*$ has no splay-like vertices, whereas $H$ has $6$ splay-like vertices. Numerically, we find that an index-$m$ solution in the $H$ class has $(m - 8)$ bend-like vertices, e.g. the index-$8$ $H$ solution has no bend-like vertices whereas the index-$14$ $H*$ solution has $6$ bend-like vertices. 
Similar remarks apply to the saddle points in the $TD$-class i.e. a $TD$-type saddle point with $m$ bend-like vertices is index-$(m+3)$.

\begin{figure}
\begin{center}
\includegraphics[width=\columnwidth]{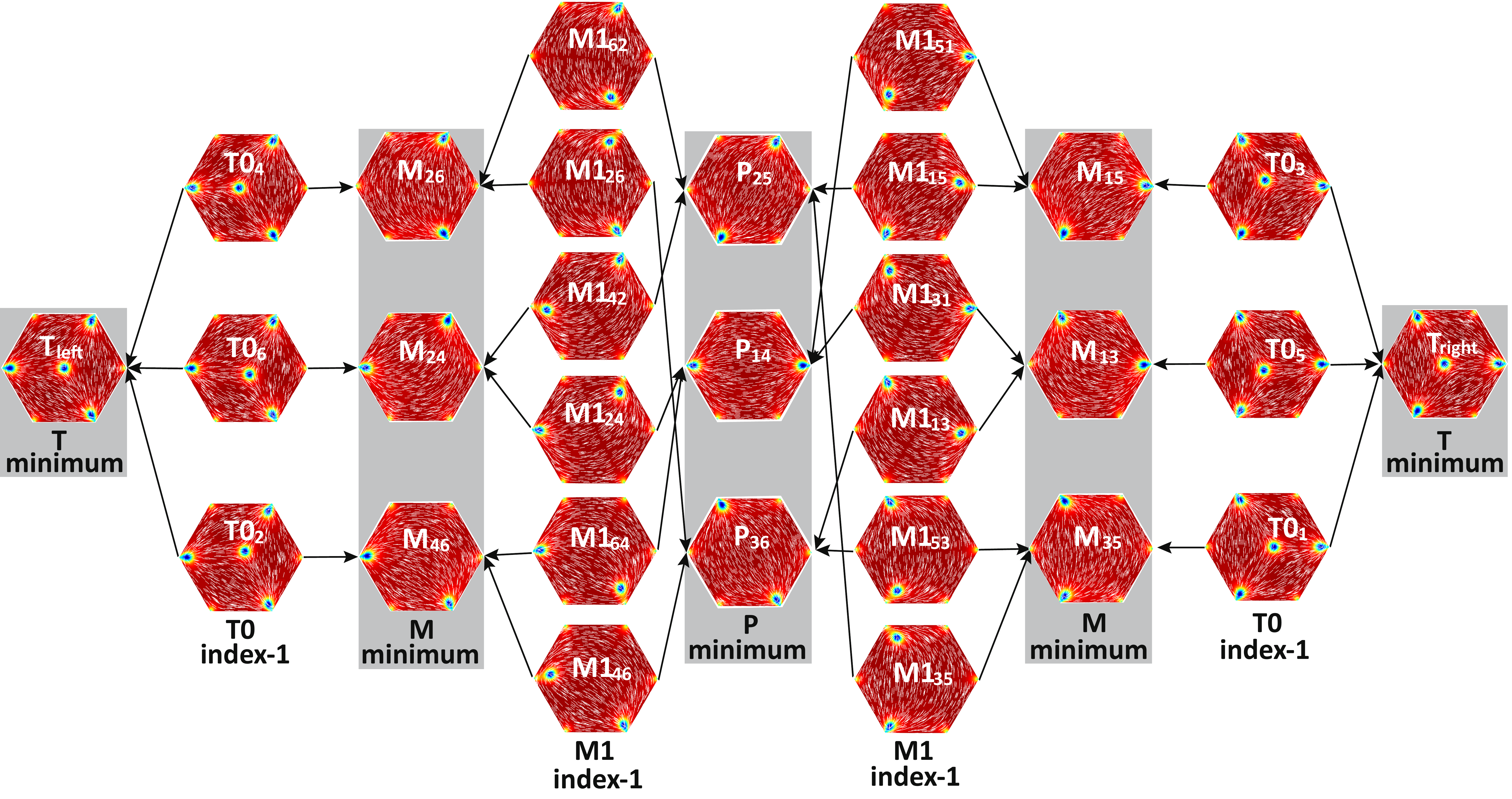}
\caption{The transition pathways between stable states including two $T$, six $M$ and three $P$ solutions, for $\bar{\lambda}^2 = 600$. Reproduced from \cite{han2020SL} with permission from IOP Publishing and the London Mathematical Society}
\label{600_transition_pathway}
\end{center}
\end{figure}

%\begin{figure}
%\begin{center}
%\includegraphics[width=\columnwidth]{./image/TD_class.pdf}
%\caption{(a) Solution landscape of the $TD$ class.
% (b) The corresponding configurations of the $TD$ class and plots of $|\P-\P^{TD}|$, where $\P$ is any solution in the $TD$ class, and $\P^{TD}$ is the index-$3$ $TD$ solution. Reproduced from \cite{han2020SL} with permission from IOP Publishing and the London Mathematical Society}
%\label{figure:TD}
%\end{center}
%\end{figure}
Next, we illustrate a comprehensive network of transition pathways between stable states including two $T$, six $M$ and three $P$ solutions, for $\bar{\lambda}^2 = 600$ in Fig. \ref{600_transition_pathway}. 
Firstly, we remark that some stable and configurationally-close solutions can be connected by a single transition state (index-$1$ saddle point) in Fig. \ref{600_transition_pathway}.
For example, the transition state between T$_{\mathrm{left}}$ and M$_{26}$ is T0$_{4}$ and the transition state between M$_{26}$ and P$_{25}$ is M1$_{62}$. 
However, two different $M$ or $P$ solutions cannot be connected by means of a single index-$1$ transition state, i.e., the transition pathway typically involves an intermediate stable $P$ or  $M$ state, risking entrapment.

The most complicated transition pathway appears to be the pathway between the two stable $T$ solutions: $T_{\mathrm{left}}$ and $T_{\mathrm{right}}$.
In fact, one numerically computed transition pathway between $T_{\mathrm{left}}$ and $T_{\mathrm{right}}$ is $T_{\mathrm{left}}$--$T0_{4}$--$M_{26}$--$M1_{62}$--$P_{25}$--$M1_{15}$--$M_{15}$--$T0_{3}$--$T_{\mathrm{right}}$, where $T0_{4}$, $M1_{62}$, $M1_{15}$ and $T0_{3}$ are transition states (index-$1$ saddle points).
This shows that a transition between two energetically-close but configurationally-far $T$ solutions may have to overcome four energy barriers and could be easily trapped by the stable $M$ or $P$ solutions. This is not a reliable way of achieving switching because of the intermediate stable states. 

\begin{figure}
\begin{center}
\includegraphics[width=\columnwidth]{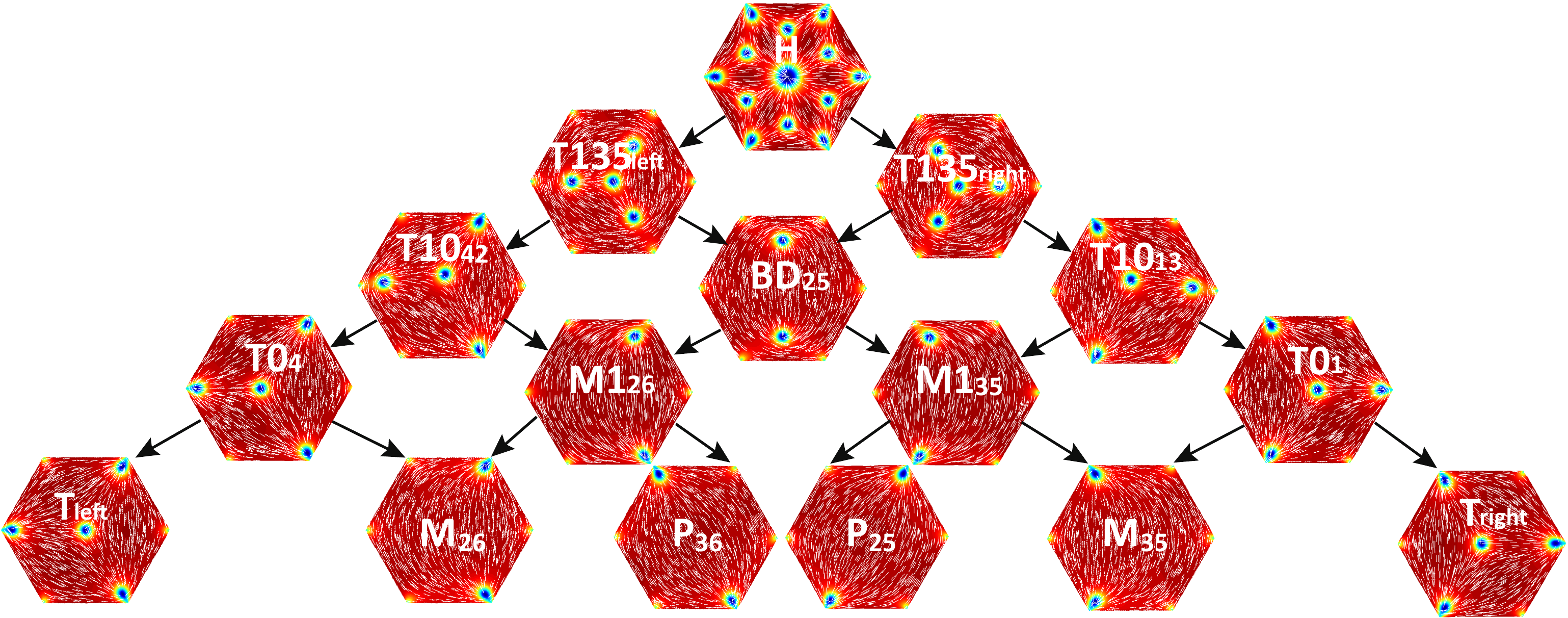}
\caption{Solution landscape starting from the $H$ solution. 
All local minima such as $T_{\mathrm{left}}$, $P_{36}$, $M_{26}$, $M_{35}$, $P_{25}$, and $T_{\mathbf{right}}$ are connected by the index-$8$ $H$ solution. Reproduced from \cite{han2020SL} with permission from IOP Publishing and the London Mathematical Society}
\label{radar}
\end{center}
\end{figure}

An alternative approach is to use higher-index saddle points with multiple unstable directions, to connect configurationally-far stable solutions. 
%The multiple unstable directions give us greater control on the dynamical pathways and offer diverse possibilities, all of which give greater insights into the design and control of solution landscapes.
Fig. \ref{radar} shows how the different $P$, $M$, and $T$ solutions are connected by high-index saddle points.
Two $M$ solutions or two $P$ solutions can be connected by the index-$2$ $BD$ solution, and the system will not be trapped by the transient local minima along this pathway.
%The benefit of this pathway mediated by a high-index saddle point as opposed to a pathway with an intermediate stable state is that the system will not be trapped by the transient local minima along this pathway. 
%Similarly, the $M$ (or $P$) solutions and $T$ solutions are connected by the index-$2$ $T10$ solution.
The $T_{\mathrm{left}}$ and $T_{\mathrm{right}}$ solutions are configurationally far and can be connected by an index-$8$ $H$ solution: $T_{\mathrm{left}}$$\leftarrow$$T135$$_{\mathrm{left}}$$\leftarrow$$H$$\rightarrow$$T135_{\mathrm{right}}$$\rightarrow$$T_{\mathrm{right}}$. The index-$8$ $H$ saddle point is connected to every stable solution and we can thus construct dynamical pathways from the $H$-solution to every individual stable solution.

%The index-$8$ $H$ solution is the stationary point in the intersection of the smallest closures of two $T$, three $P$ and six $M$ solutions on the energy landscape.
%The index-$8$ $H$ saddle point is connected to every stable solution and we can thus construct dynamical pathways from the $H$-solution to every individual stable solution.

Our numerical results highlight the differences between transition pathways mediated by index-$1$ saddle points and pathways mediated by high-index saddle points.
We deduce that index-$1$ saddle points are efficient for connecting configurationally-close stable solutions. 
For configurationally-far stable states, they are generally connected by multiple transition states and intermediate stable states, or it may be possible to find a dynamical pathway between these configurationally-far stable states via high-index saddle points. The selection of dynamical pathways is an open problem of tremendous scientific and practical interest.
%we can construct transition pathTays composed of transition states and intermediate stable states but these pathways have multiple s.pdf and the system could easily be trapped in an intermediate state. A high-index saddle point would expedite the switching with fewer s.pdf and no intermediate stable states, but we expect the energy barrier to be significantly higher. We plan to make quantitative studies on these lines in the future.

\begin{figure}[b]

\includegraphics[width=0.649\columnwidth]{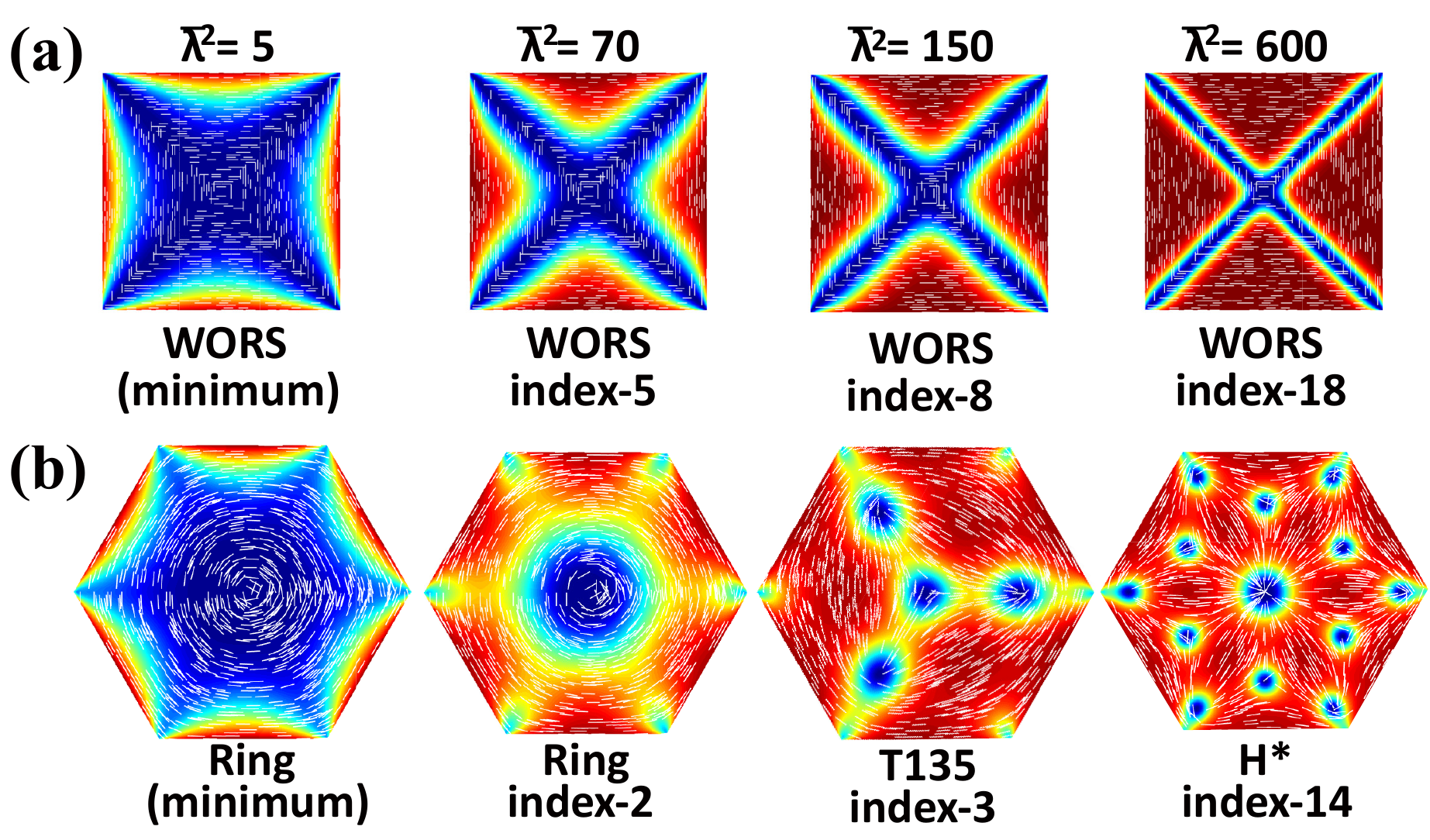}
\caption{Comparison of the parent states of the solution landscapes on the square (a) and the hexagon (b). On square, the parent state is always $WORS$. Whilst, on hexagon, the parent state change from $Ring$, $T135$, to $H$ state. Reproduced from \cite{han2020SL} with permission from IOP Publishing and the London Mathematical Society}
\label{comparison}
\end{figure}

Finally, let us compare the solution landscapes on a hexagon with that on a square domain. This illustrates the effects of geometry on solution landscapes. The most obvious difference is on the parent state.
%In Fig. \ref{comparison} (a), 
The Morse index of the $WORS$ increases with the domain size, $\lambda$ and the $WORS$ is always the parent state for a square domain \cite{yin2020construction}.
Intuitively, this is because the diagonal defect lines become longer, so that the Morse index of the $WORS$ also increases with increasing edge length/increasing $\lambda$.
%\hl{Whilst, the $Ring$  solution on a hexagon, the analogue of WORS, has the dimension-$0$ point defect and maintains index-$2$ after the bifurcation from the index-$0$ $Ring$  solution to the index-$2$ $Ring$  solution./
The $Ring$  solution, which is the analogue of the $WORS$ on a hexagon, is index-$0$ for $\lambda$ small enough, and is an index-$2$ saddle point solution for larger $\lambda$, i.e., the Morse index does not increase with increasing $\lambda$.
The highest-index parent saddle point on a hexagon changes from the $Ring$  solution to the index-$3$ $T135$ and index-$14$ $H*$ (see Fig. \ref{comparison} (b)) respectively, where $T135$ and $H*$ solutions emerge through saddle-node bifurcations, as $\lambda$ increases. 
We believe that the hexagon is a more generic example of a regular polygon with an even number of sides than a square, and hence, we expect that the qualitative aspects of our numerical study on a hexagon will extend to arbitrary polygons with an even number of sides.

\section{Conclusions and Discussions}\label{sec:conclusion}
This review focus on NLC equilibria in reduced 2D settings, within the reduced LdG framework (\ref{eq:reduced}). We look at regular polygons, and the effects of elastic anisotropy captured by a parameter $\hat{L}_2$, with some preliminary work on the effects of geometrical anisotropy. The geometrical size is captured by a typical length (e.g. edge length of a polygon), denoted by $\lambda$. The $\lambda \to 0$ and $\lambda \to\infty$ limits are analytically tractable. In fact, we have a unique globally stable NLC equilibrium for $\lambda$ sufficiently small, and multistability for $\lambda$ sufficiently large. The shape of the geometry plays a crucial role in the structural details. For example, the $WORS$ with a pair of mutually orthogonal defect lines along the diagonals is exclusive to a square domain, where we observe the $Ring$-solution with a central $+1$-defect for all other polygons except the equilateral triangle. For a regular triangle, the stable NLC equilibrium has a central $-1/2$-defect in the $\lambda \to 0$ limit. For a $K$-regular polygon with $K$ edges, there are at least $\left[\frac{K}{2}\right]$-classes of stable NLC equilibria for $\lambda$ large enough, so that the shape of the polygon has a crucial role in multistability.

The effects of $\hat{L}_2$ have only been reviewed on square domains. Elastic anisotropy destroys the perfect $WORS$-cross structure for small $\lambda$, enhances the stability of some symmetric critical points and very importantly, introduces a novel $Constant$-solution branch for large values of $\hat{L}_2$. The square is special, and we need more comprehensive studies on generic 2D domains to capture the effects of $\hat{L}_2$ on solution landscapes.

Of particular interest are the study of saddle points and dynamical pathways between NLC equilibria on a 2D hexagon, at a fixed temperature below the nematic supercooling temperature. We review results on high-index saddle points from \cite{han2020SL}, focusing on the effects of $\lambda$, and numerically illustrate dynamical pathways, with intermediate index-$1$ saddle points/transition states versus dynamical pathways with intermediate high-index saddle points. The high-index saddle points are poorly understood in the literature but can play a crucial role in switching, selection of stable states and transient non-equilibrium dynamics, all of which are relevant to applications of confined NLC systems.

With regards to future research avenues, the possibilities are tremendous. A natural question concerns the sensitivity of solution landscapes to shape variations i.e. if the geometry is not fixed but can be optimised with regards to prescribed properties. In other words, can we use shape and topology to tune the Morse indices of critical points in the LdG framework? Similarly, can we mathematically analyse new composite materials with multiple order parameters e.g. a nematic order parameter and a magnetic order parameter; see \cite{hanpre2021} for detailed numerical studies of a prototype model for ferronematics, on 2D polygons with tangent boundary conditions. Last but not the least, these problems rely on a delicate and challenging combination of tools from variational analysis, numerical analysis, simulations, and experiments. In \cite{maity2021discontinuous} and \cite{maity2021error}, the authors perform numerical analyses of some finite-element methods for the reduced LdG model in \eqref{eq:reduced}, with \emph{a priori} and \emph{a posteriori} estimates for the Discontinuous Galerkin Method and the Nitsche's method. Of course, the possibilities are endless and our vision is to design and implement generic algorithms for partially ordered materials, that can select the best mathematical model for the system under consideration, and then do comprehensive searches of the solution landscapes, yielding deterministic recipes for NLC-based systems which are  predicted, designed and controlled by mathematical toolboxes.

\section*{Acknowledgments}
AM gratefully acknowledges support from the University of Strathclyde New Professor Fund, a Leverhulme International Academic Fellowship, a Royal Society Newton Advanced Fellowship and the DST-UKIERI grant on "Theoretical and experimental studies of suspensions of magnetic nanoparticles, their applications and generalizations". AM gratefully acknowledges support from an OCIAM Visiting Fellowship, a Visiting Professorship from the University of Bath and a Visiting Professorship from IIT Bombay (India). YH is fully supported by a Royal Society Newton International Fellowship. AM and YH are grateful to Professor Lei Zhang (Peking University), Professor Lei Zhang (Shanghai Jiao Tong University) and Dr Lidong Fang for helpful suggestions.

\section{Supplement: Numerical Methods}\label{sec:num}
We have used various methods to discretize the domain $\Omega$, in the case studies of this review. In Section \ref{sec:pol} and \ref{sec:ani}, on arbitrary regular polygons, we use standard Finite Element Methods to solve the linear systems including the Laplace equation and the limiting problems in the $\lambda \to\infty$ limits. %, by using LU solver and combine the Newton's method to solve the nonlinear system of EL equations. 
All finite-element simulations and numerical integrations are performed using the open-source package FEniCS~\cite{olgg2012fenics}, along with the LU solver and the Newton's method. %The solutions of EL equations are the critical points of the corresponding free-energy functional. 
Newton's method strongly depends on the initial condition. We typically use the analytic solutions in the asymptotic limits --- e.g. the $\lambda \to 0$ or $\lambda \to \infty$ limit and the $\hat{L}_2\to 0$ limit, or perturbations of these solutions, as the initial conditions for the numerical solver, for a range of values of $\lambda$. 
In Section \ref{sec:ani}, on square domain, we use traditional finite difference schemes for square mesh. In Section \ref{sec:SL}, on hexagon domain, we apply finite difference schemes over triangular elements to approximate the spatial derivatives, by analogy with the conventional discretization of a square domain \cite{fabero2001explicit}. 

To compute the bifurcation diagrams consisting of known stable and unstable solution branches, we perform an increasing $\lambda$ sweep for the unique solution branch such as $WORS$, $Ring$ , $Constant$ for small $\lambda$, and decreasing $\lambda$ sweep for the distinct $Para$, $Meta$, $Ortho$, $D$ or $R$ solution branches. We distinguish between the distinct solution branches by defining two new measures, e.g. $\int_{\Omega} P_{12}\left(1+x+y\right)dxdy$ and $\int_{\Omega} P_{11}\left(1+x+y\right)dxdy$, and plot these measures versus $\lambda^2$ for the different solutions. Actually, the specific form of measure depends on the central point and the shape of the domain $\Omega$, and the mathematical model.
We study the stability of the solutions by numerically calculating the smallest real eigenvalue of the Hessian of the free energy and the corresponding eigenfunction using the LOBPCG (locally optimal block preconditioned conjugate gradient) method in \cite{yin2019high} (which is an iterative algorithm to find the smallest (largest) $k$ eigenvalues of a real symmetric matrix.) A negative eigenvalue is a signature of instability and we have local stability if all eigenvalues are positive.

To investigate unstable solutions of the Euler--Lagrange equations, labelled as saddle points, in Section \ref{sec:SL}, we use the high-index optimization-based shrinking dimer (HiOSD) method to compute any-index saddle points \cite{yin2019high}. The high-index saddle dynamics for finding an index-$k$ saddle point can be viewed as a transformed gradient flow for the state variable $\x$ and $k$ direction variables $\mathbf{v}_i$. The stability analysis is performed to show that a linearly stable steady state of this dynamical system is exactly an index$-k$ saddle point.% and $k$ eigenvectors of the Hessian at the index-$k$ saddle point corresponding to the $k$ negative eigenvalues. 
The HiOSD method is an efficient tool for the computation of unstable saddle points and (local and global) minimizers, without good initial guesses. The connectivity of saddle points, including transition pathways, can be well established via the downward search and upward search algorithms. By combining the HiOSD method with downward and upward search algorithms, we can construct the solution landscape  systematically. For more details, the readers are referred to the reference \cite{yin2020construction}. In general, %in the study of solution landscape 
we track bifurcations by tracking the indices of solutions; a change in the index is a signature of a bifurcation and a possible change of stability properties.

\bibliographystyle{unsrt}
\bibliography{WMR}
\end{document}